\documentclass{elsart}
\usepackage{graphicx}
\usepackage{amssymb}

\begin{document}

\begin{frontmatter}
\title{Surface anisotropy in nanomagnets: Transverse or N\'eel~?}
\author{H. Kachkachi}
\ead{Corresponding author: kachkach@physique.uvsq.fr}
\author{and H. Mahboub\thanksref{$}}
\thanks[$]{On leave of absence from: Laboratoire de Physique du Solide,
Facult\'e des Sciences Dhar El Mehraz, B.P. 1796 Atlas, F\`es, Morocco.}
\address{Laboratoire de Magn\'{e}tisme et d'Optique,
Universit\'e de Versailles St. Quentin, \\
45 av. des Etats-Unis, 78035 Versailles, France}

\begin{abstract}
Through the hysteresis loop and magnetization spatial distribution we
study and compare two models for surface anisotropy in nanomagnets:
a model with transverse anisotropy axes and N\'eel's model.
While surface anisotropy in the transverse model induces several jumps in the
hysteresis loop because of the cluster-wise switching of spins, in the N\'eel
model the jumps correspond to successive {\it coherent partial rotations}
of the whole bunch of spins.
These calculations, together with some hints from available experimental
results, suggest that N\'eel's model for surface anisotropy is more appropriate.
\end{abstract}

\end{frontmatter}

\section{\label{intro}Introduction}
Understanding surface effects in magnetic nanoparticles is a big challenge from
both the experimental and theoretical points of view.
Experimentally, this challenge resides mainly in the fact
that the current experimental techniques do not allow for probing the
microscopic structure of the surface in round geometries, as is the case for
$2d$ magnetism~\cite{frietal94jmmm}.
In particular, we so far have no clear idea as to what kind of surface
anisotropy we have in a nanoparticle.
The corresponding constant has been estimated by many authors in an indirect way
from neutron scattering experiments \cite{gazetal97epl}, magnetic measurements
\cite{doretal97acp}, and FMR experiments~\cite{shiphd99}, by fitting the
results assuming an effective anisotropy constant.
This is a rough estimation that is only valid when deviations from collinearity
are very small.
On the other hand, the easy direction of the magnetization on
the surface, let aside that of the atomic magnetic moments, is still unknown.
In fact, even the anisotropy in the core of the particle is assumed to be that
of the bulk material.
As such, to investigate surface effects on the properties of nanoparticles
theorists can only ``try" some models.
The most often used model is what we call here {\it transverse surface
anisotropy} (TSA) model which attributes an easy axis along the transverse
(radial) direction to each spin on the boundary [see the textbook~\cite{aha96}
and references therein].
Another more physically plausible model is what we call {\it N\'eel surface
anisotropy} (NSA) model, developed by N\'eel~\cite{nee53cras54jpr}. This
model is more realistic because the anisotropy at a given lattice site only
occurs if the site's environment presents some defects, e.g., in the
coordination number. %
In \cite{kacdim02prb} (see also \cite{dimwys94prb}) we used the TSA
model and studied the details of the switching process in a spherical particle
when the surface anisotropy constant $K_s$ is varied.
We solved the coupled Landau-Lifshitz equations with local and global rotation
constraints.
In particular, we found that there is a characteristic value of $K_s$ separating
two different regimes: For small $K_s$, as compared to the exchange
coupling, the switching is basically Stoner-Wohlfarth like, i.e., coherent,
whereas for large values the magnetization switching can no longer be regarded
as coherent, as it operates cluster-wise.
In \cite{garkac03prl} we used the Green's function technique in the continuum
limit, in the absence of magnetic field, and computed for
the NSA model the contribution of the surface to the anisotropy energy of a
spherical particle with simple cubic lattice structure.
We found that this contribution is second order in the surface anisotropy
constant $K_s$ [see Eq.~(\ref{En_greenfunct}) below], scales with the 
particle's volume ${\mathcal N}$ and has cubic symmetry with preferred
directions $\left[\pm,\pm,\pm\right]$.

The present work follows on the previous ones~\cite{kacdim02prb},
\cite{garkac03prl} and is an attempt to give some hints for answering the
question addressed in the title, namely which one of these models is most
adequate to describe surface effects in a small magnetic system.
We have no pretention here to give a definite answer to this question as we
believe that only with the help of experiments, e.g., measuring the
hysteresis cycle of an isolated nanoparticle, could one do so.
However, today technical difficulties prevent such measurements [see the
review \cite{wer01acp}].
Meanwhile, it is important to understand the effects of surface
anisotropy within the framework of the current models, and eventually
make some predictions for future investigations.
In addition, we do not try to be exhaustive since a detailed study
of the magnetization switching within the TSA model has already been presented
in \cite{kacdim02prb}.
Instead, we only investigate the main difference between the two models with
regard to the hysteresis loop and magnetic structure so as to gain some insights
into their relevance to the study of surface effects in nanomagnets.
More precisely, we compute the hysteresis loop and magnetic structures of a
spherical particle with simple cubic lattice structure and uniaxial anisotropy
in the core with easy axis $z$.
We consider both cases of vanishing and non-vanishing exchange interaction
between spins inside the particle.
The unrealistic non-interacting case is considered only because it allows for
a comparison of the TSA and NSA in pure form.
\section{Statement of the problem}
\label{sec:statement}
Our model Hamiltonian is given by the (classical)
anisotropic Dirac-Heisenberg model~\cite{kacdim02prb}

\begin{equation}\label{DH}
H = -\sum\limits_{\left\langle i,j\right\rangle }J_{ij}{\bf
s}_{i}\cdot {\bf s}_j - (g\mu _{B}){\bf H}\cdot\sum\limits_{i=1}^{\mathcal N}{\bf
s} _{i} + H_{an},
\end{equation}
where ${\bf s}_{i}$ is the unit spin vector on site $i$, ${\bf H}$ the
uniform magnetic field, ${\mathcal N}$ the total number of spins (core and
surface), and $J_{ij} ( = J>0)$ the nearest-neighbor exchange coupling.
$H_{an}$ can be the uniaxial single-site anisotropy energy
\begin{equation}\label{uaa}
H_{an} = -\sum\limits_{i}K_i({\bf s}_{i}\cdot{\bf e}_i)^{2},
\end{equation}
with easy axis ${\bf e}_{i}$ and constant $K_i>0 $.
If the spin at site $i$ is in the core, the anisotropy axis ${\bf e}_i$ is taken
along the reference $z$ axis and $K_i=K_c$. Otherwise, for surface spins, this
axis is along the radial (or transverse) direction and $K_i=K_s$.
In this case, the model in (\ref{uaa}) for surface spins is the TSA model.
The second model that will be studied and compared with the previous one, is
the more physically appealing model of surface anisotropy that was
introduced by N\'eel \cite{nee53cras54jpr},
\begin{equation}\label{NSA}
H_{an}^{\mbox{N\'eel}} = -K_s\sum\limits_{i}\sum\limits_{j=1}^{z_i}({
\bf s}_{i}{\bf \cdot e}_{ij})^{2},
\end{equation}
where $z_i$ is the coordination number of site $i$ and ${\bf e}_{ij}={\bf
r}_{ij}/r_{ij}$ a unit vector connecting the site $i$ to its nearest
neighbors $j$.
This model is more realistic because the anisotropy at a given
site occurs only when the latter loses some of its neighbors, i.e., when it
is located on the boundary. This is the NSA model.

In this work we ignore dipolar interactions between spins
inside the particle, since in our case of small spherical particles, the
volume term is negligible \cite{hah70prb} and the shape anisotropy yields an
irrelevant constant.
In addition, here we restrict ourselves to the investigation of ``pure" surface
anisotropy within the two models.

In the sequel, we will use the following notation $h \equiv H/2K_c$ for field,
with the magneto-crystalline ainsotropy constant $K_c$ in the core taken as the
energy scale,
$j\equiv J/K_c$ for the exchange coupling, which will be taken in
our calculations the same everywhere inside the particle,
$k_s\equiv K_s/K_c$, and $D$ denotes the particle's diameter.

We study hysteretic properties of a nanoparticle by solving the
(local) Landau-Lifshitz equations at zero temperature.
The method used to obtain the hysteresis loop and other
characteristics were given in great detail in \cite{kacdim02prb} and
will not be repeated here.
We compare the implications of the two models, TSA and NSA, on
the hysteretic properties of a nanoparticle, such as hysteresis loop and
coercive field as a function of the surface anisotropy strength $k_s$.
In all these calculations the anisotropy in the core is taken as uniaxial along
the reference axis $z$, and the field is applied at an angle of $\pi/4$ with
respect to the latter in the $x-z$ plane.
\section{Results and discussion}
\label{sec:Results and discussions}
\subsection{Noninteracting case}
\label{subsec: zer0_exchange}
\begin{figure}[h!]
\begin{center}
\includegraphics[angle= -90, width= 14cm]{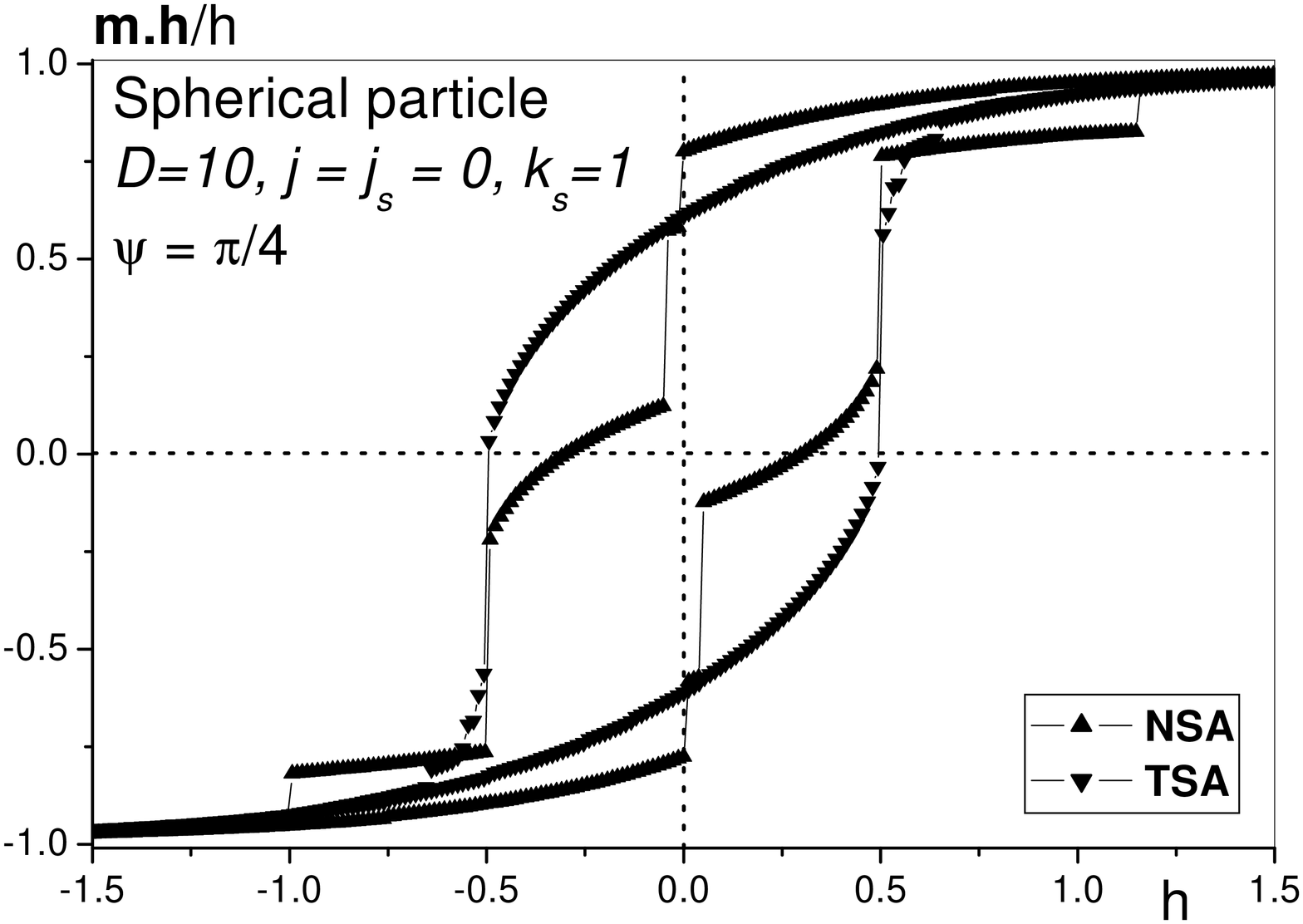}
\end{center}
\caption{\label{T_N_j0_D10_Ks1}
Hysteresis loops for a spherical particle with $D = 10~({\mathcal N}=360), j = 0,
k_s = 1$ for TSA and NSA models.}
\end{figure}

\begin{figure}[h!]
\begin{center}
\includegraphics[angle=-90,width=3cm]{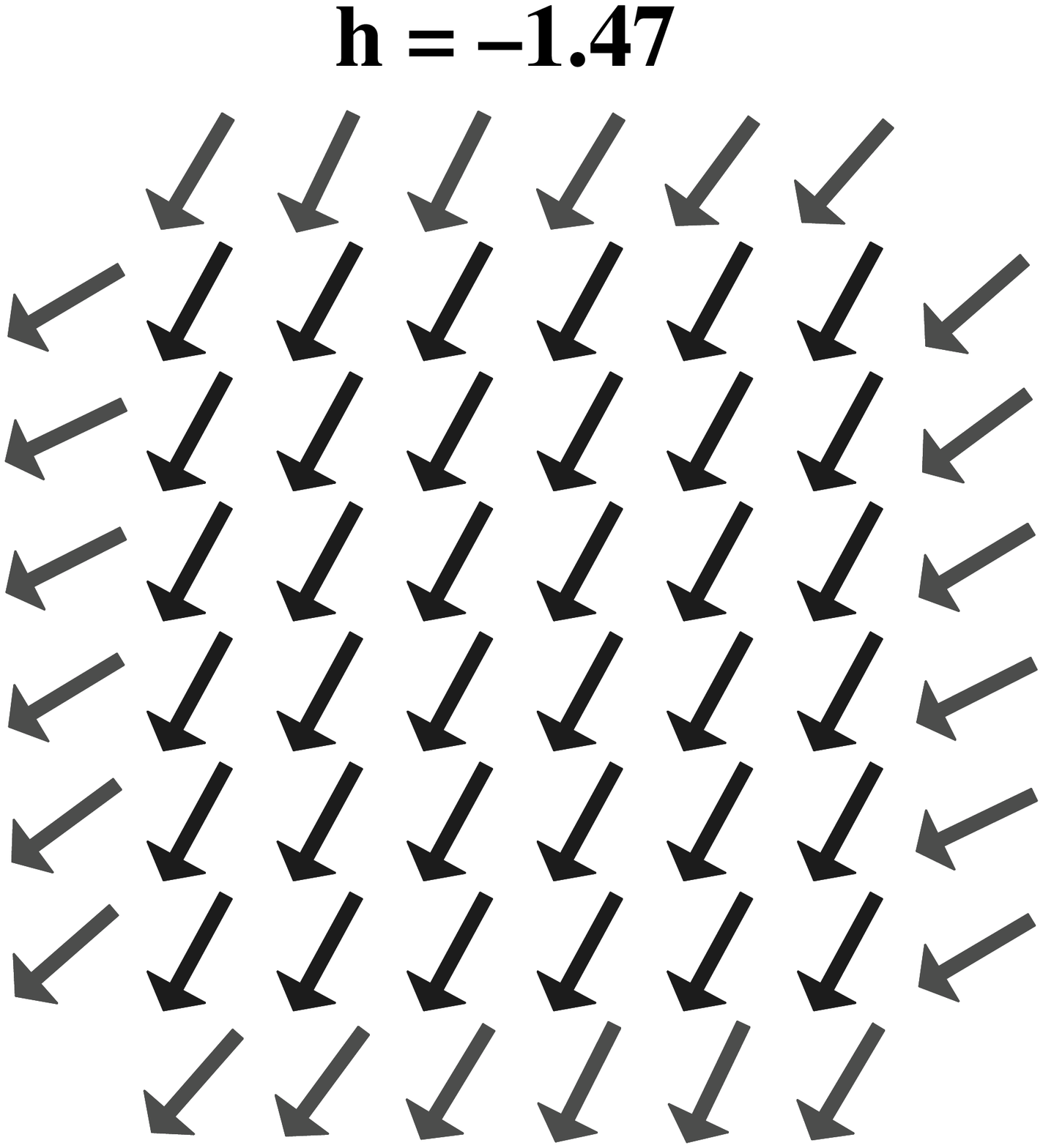}\hspace{0.5cm}
\includegraphics[angle=-90,width=3cm]{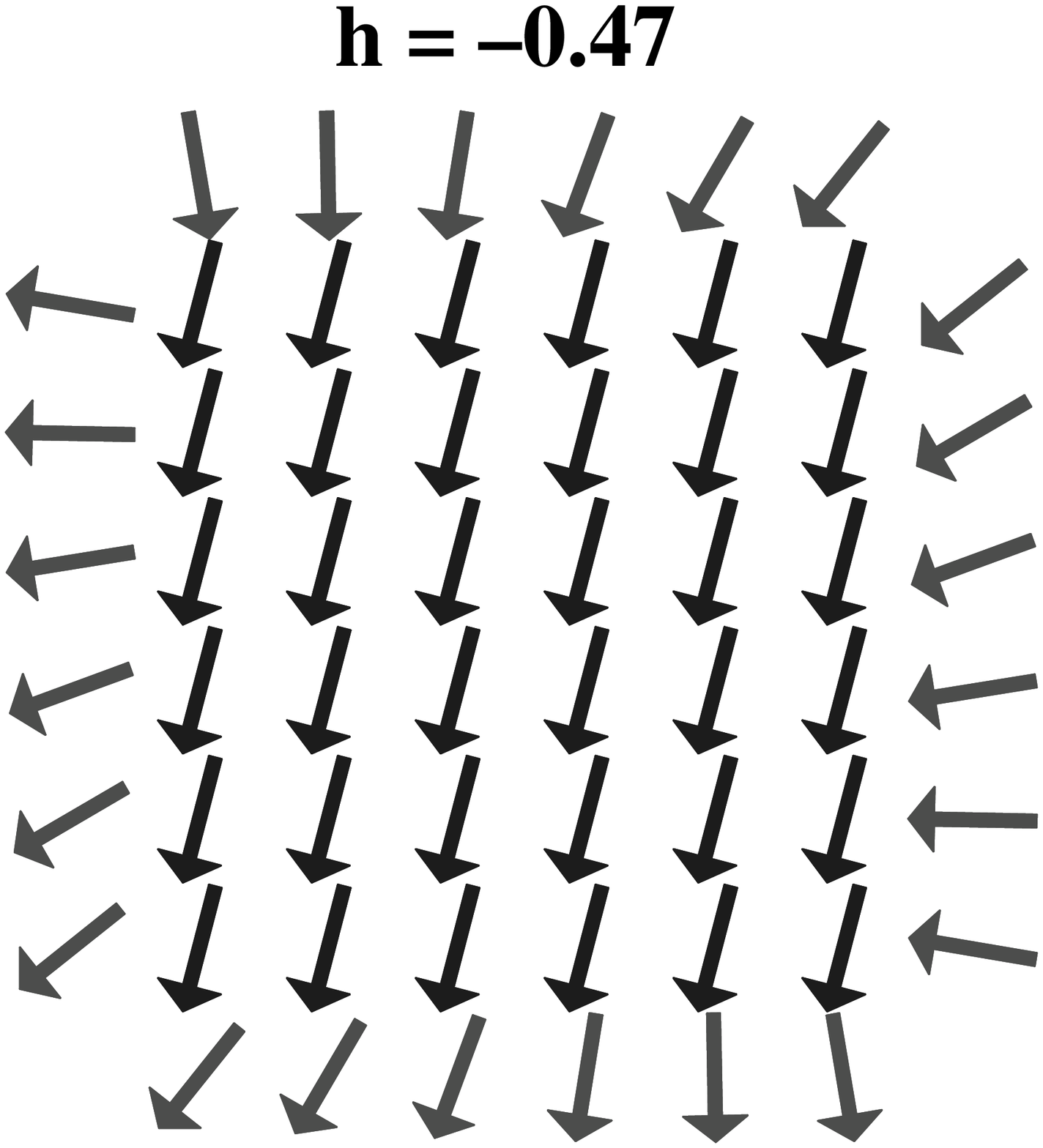}\hspace{0.5cm}
\includegraphics[angle=-90,width=3cm]{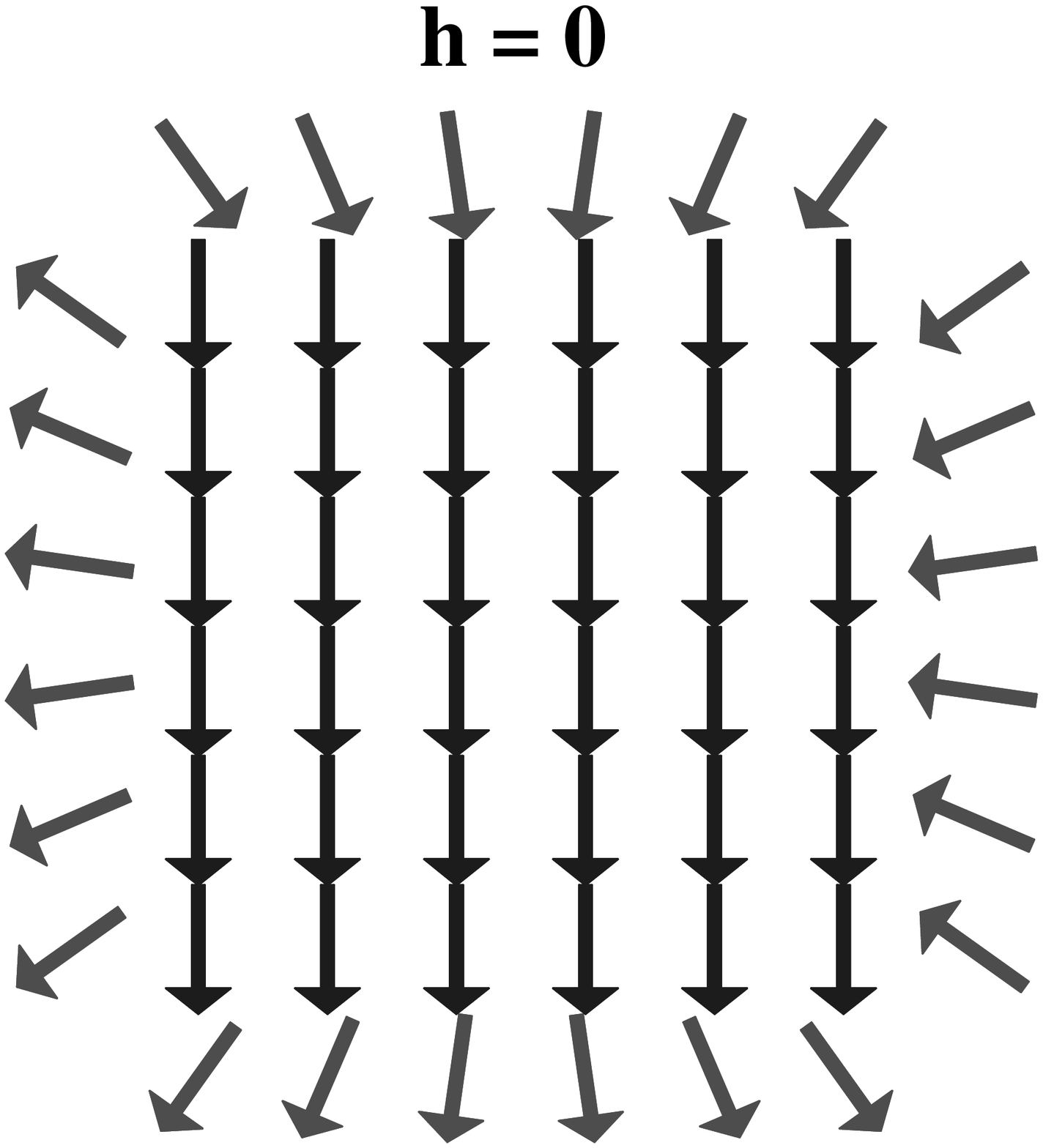}\hspace{0.5cm}
\includegraphics[angle=-90,width=3cm]{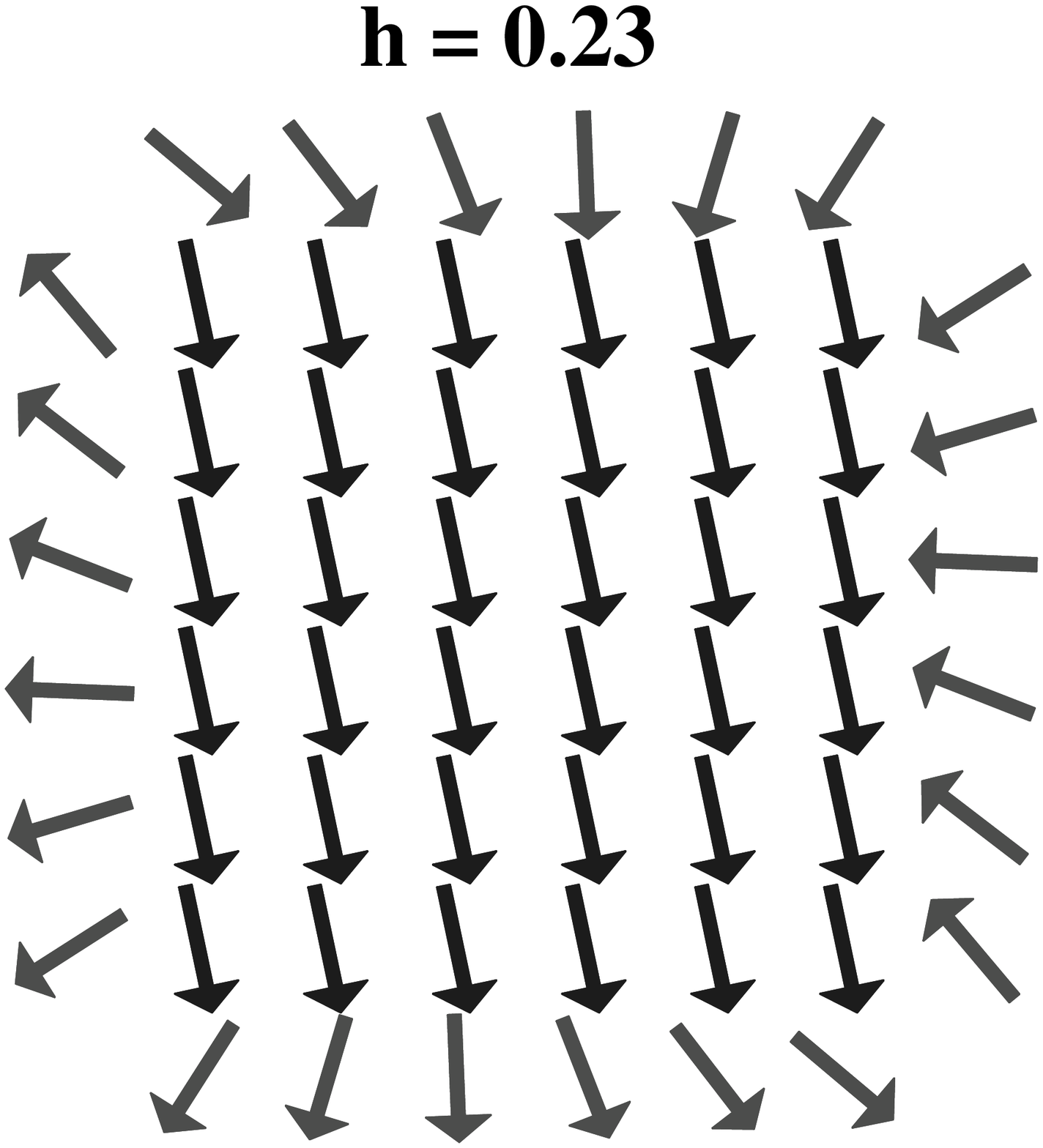}\\
\vspace{0.5cm}
\includegraphics[angle=-90,width=3cm]{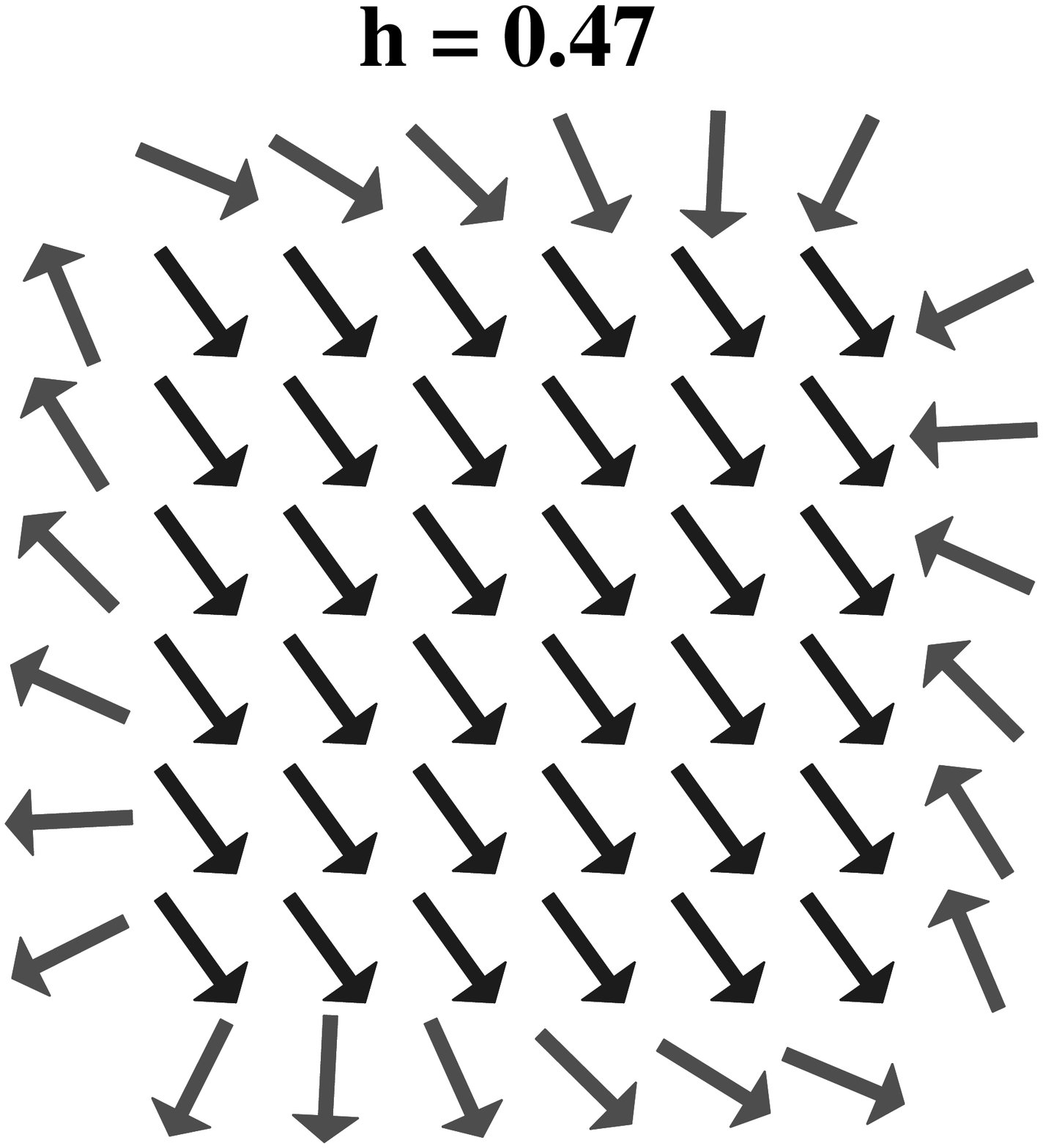}\hspace{0.5cm}
\includegraphics[angle=-90,width=3cm]{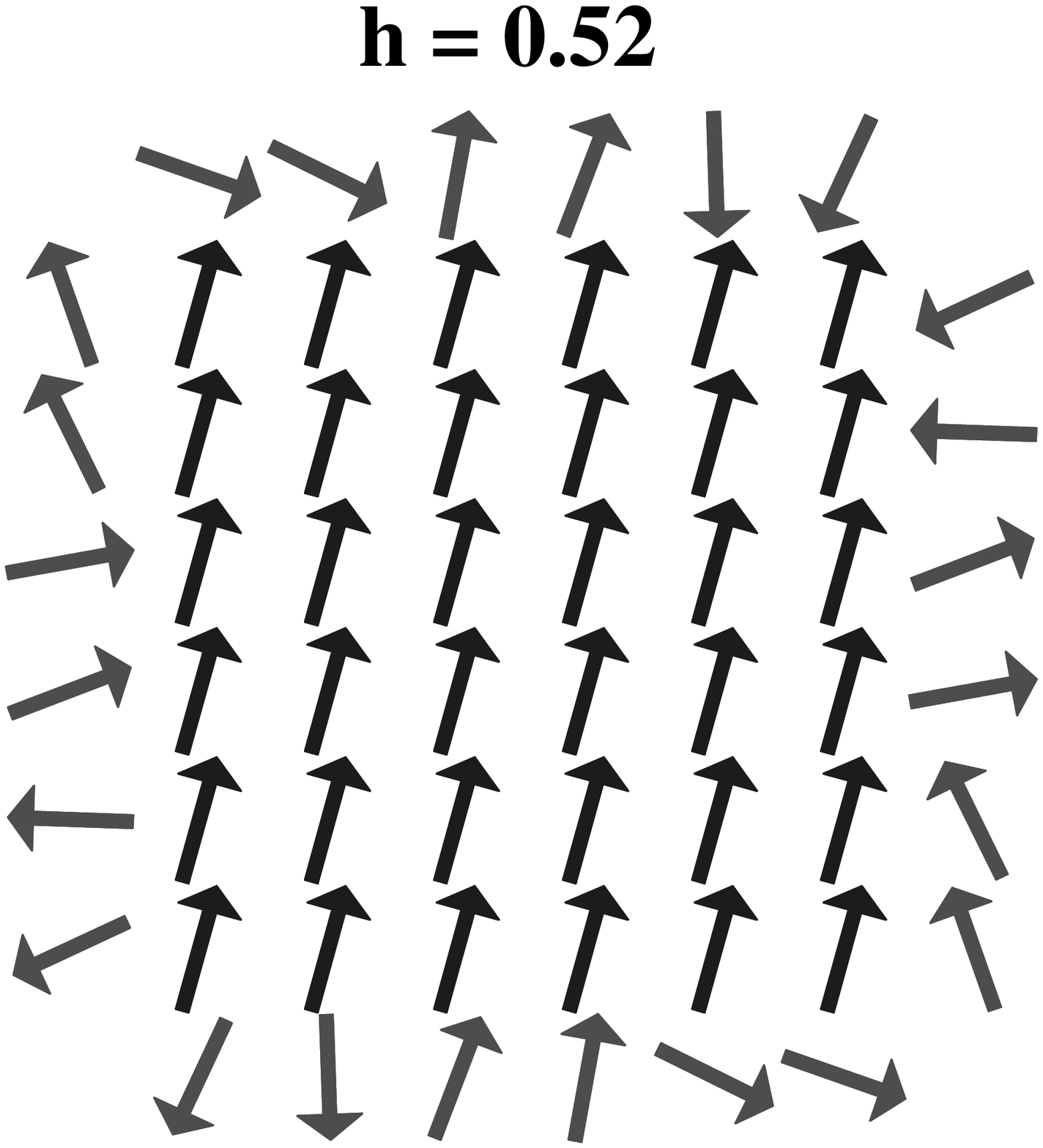}\hspace{0.5cm}
\includegraphics[angle=-90,width=3cm]{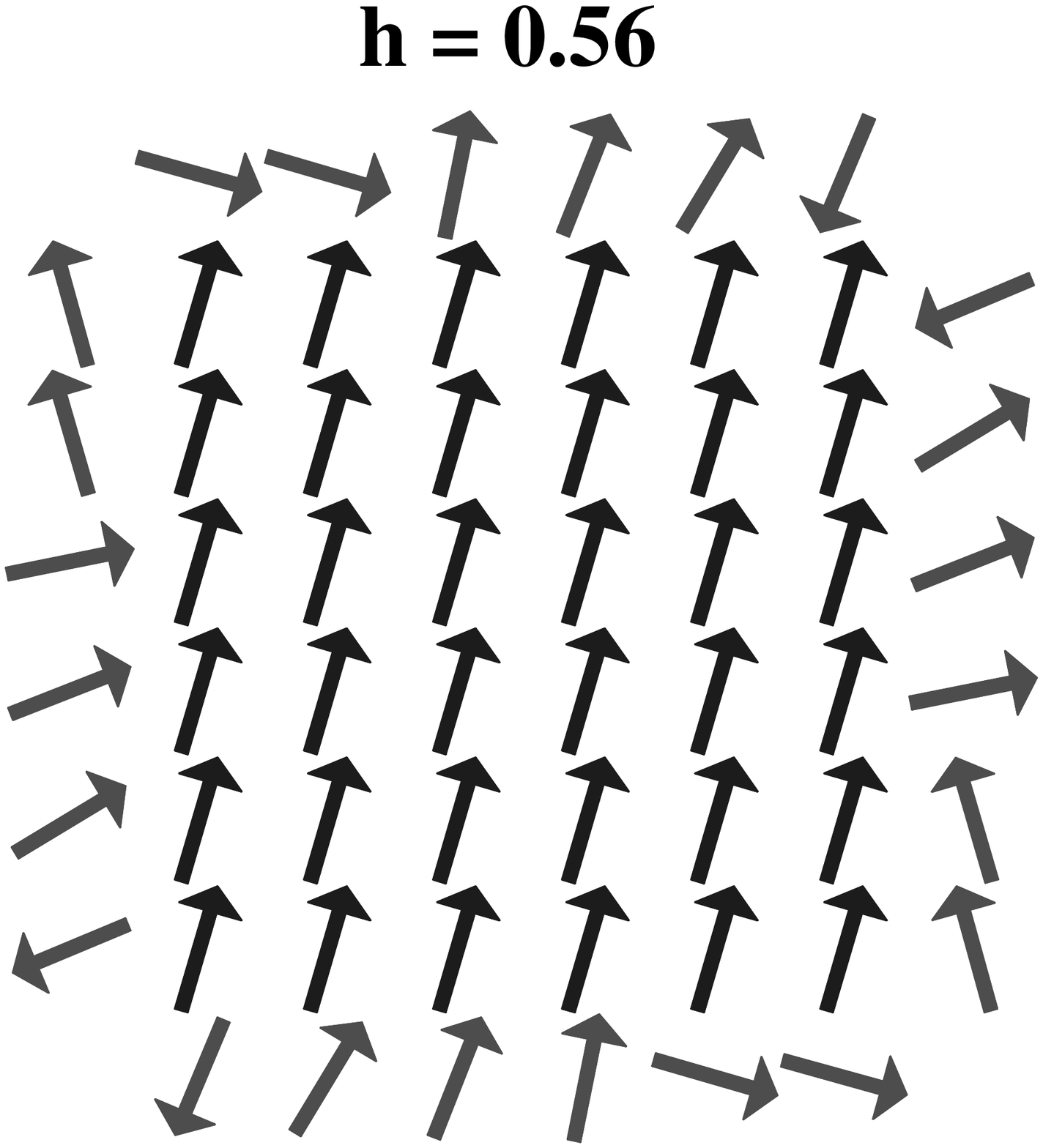}\hspace{0.5cm}
\includegraphics[angle=-90,width=3cm]{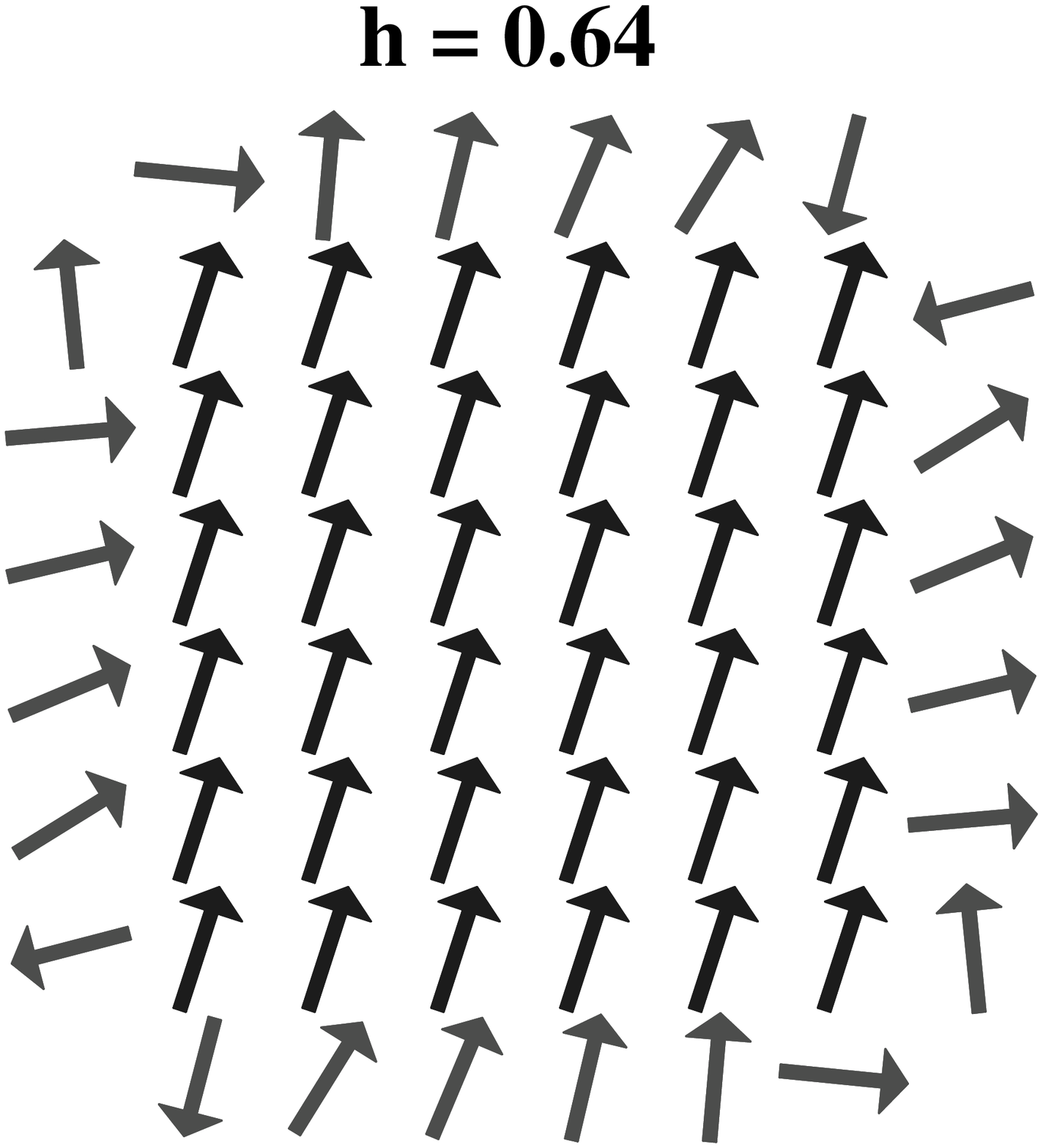}\\
\vspace{0.5cm}
\includegraphics[angle=-90,width=3cm]{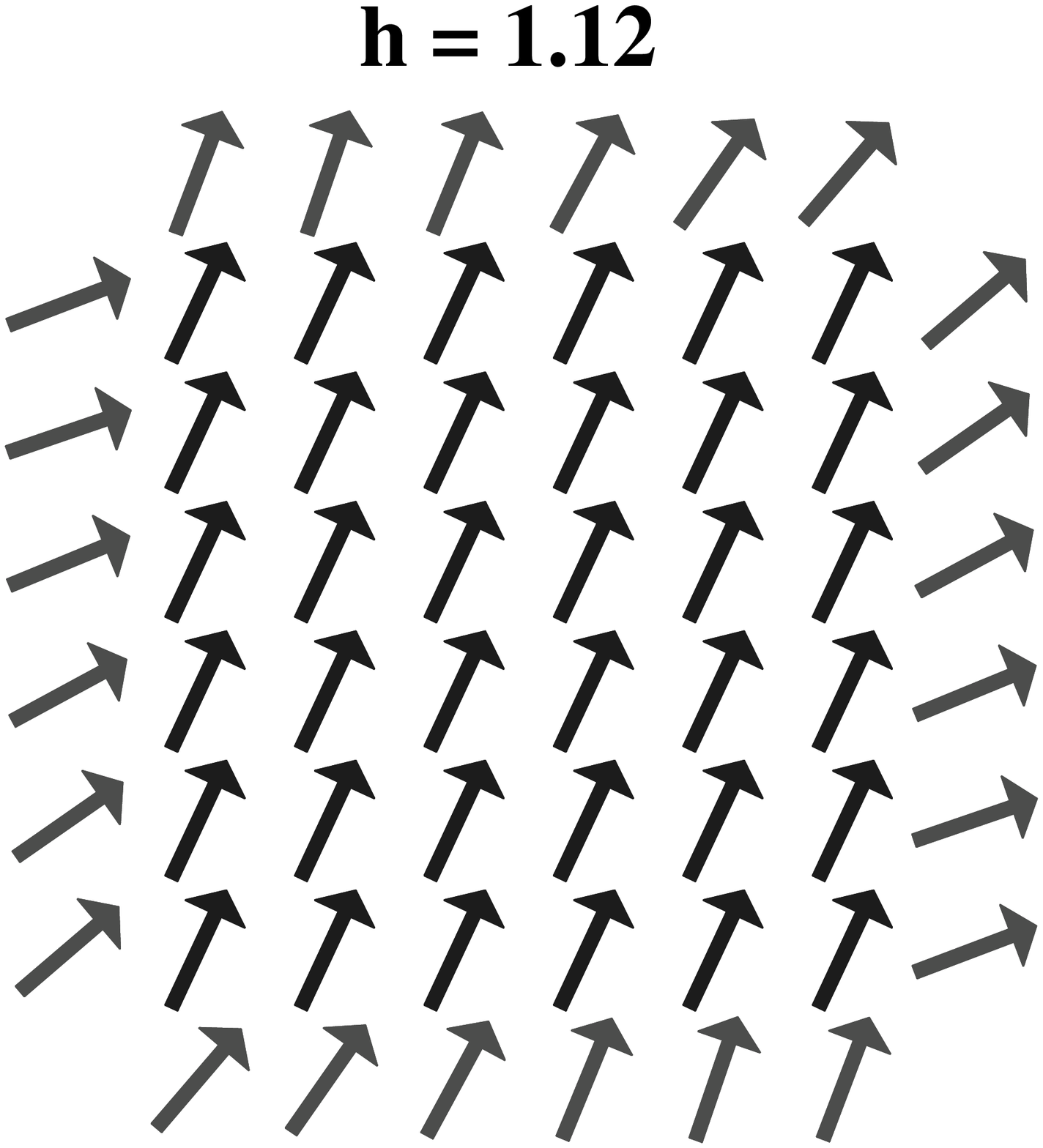}\hspace{0.5cm}
\includegraphics[angle=-90,width=3cm]{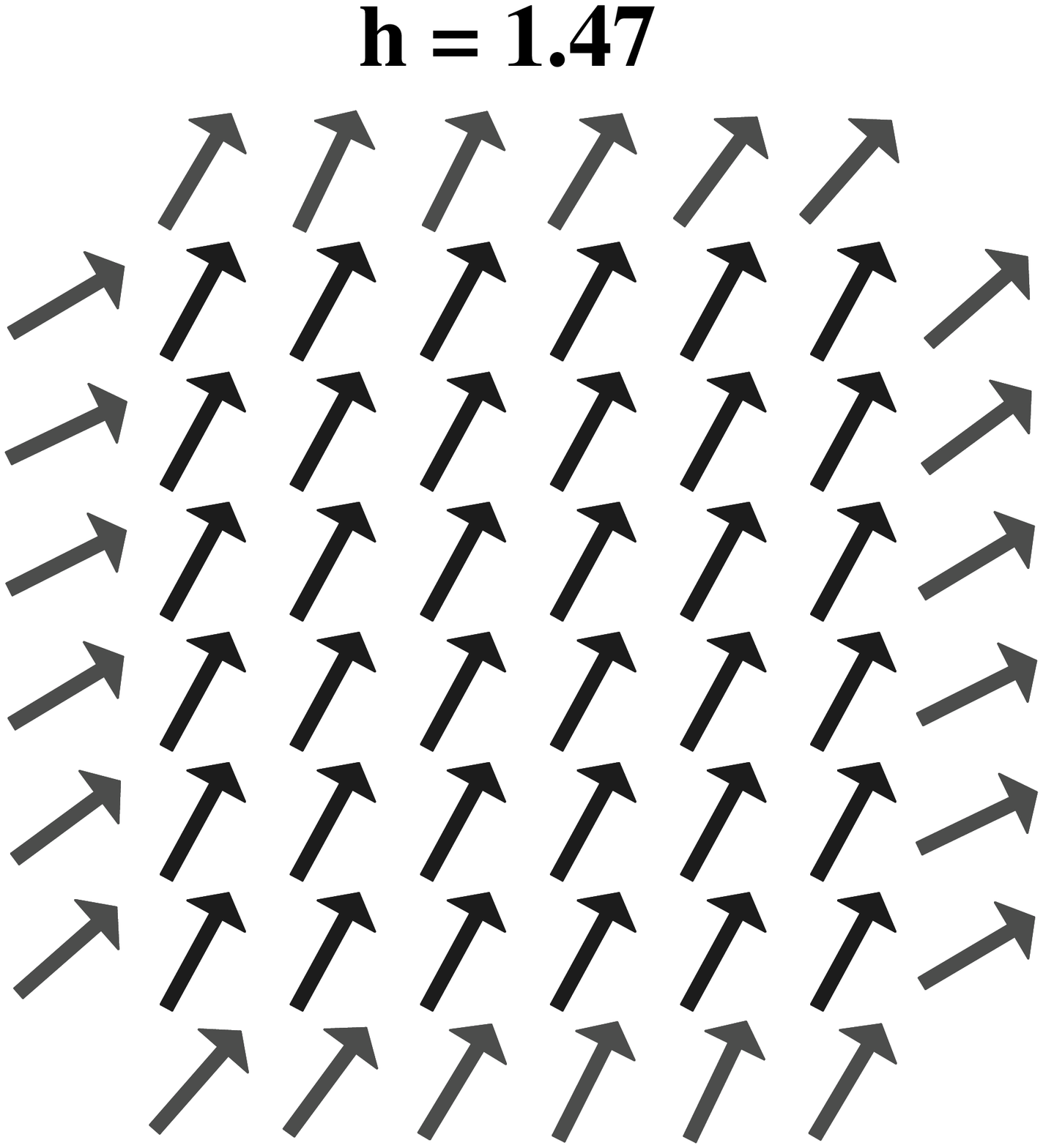}
\end{center}
\caption {\label{struct_T_D10_j0_K1}
Magnetic structures corresponding to the middle plane of the particle for the
TSA model.
The field values are indicated on top of the structures and correspond to
the ascending hysteresis-loop branch of Fig.~\ref{T_N_j0_D10_Ks1} for TSA.}
\end{figure}

In order to compare the TSA and NSA models in pure form, we first consider the
case of a spherical particle of non-interacting spins ($J_{ij}=0$), and
the core-anisotropy constant equal to that on the surface.
In Fig.~\ref{T_N_j0_D10_Ks1} we plot the hysteresis loop for both TSA and NSA
models.
%
%
First, in both cases the core switches at the same field of $1/2$ since the
anisotropy axis in the core is along $z$ and makes the angle $\pi/4$ with
respect to the field direction.
In the case of TSA, at zero field the surface spins are obviously directed along
the radial direction [see Fig.~\ref{struct_T_D10_j0_K1} $h=0$].
At non zero field, as was discussed in detail in \cite{kacdim02prb}, because
of the radial direction of the surface anisotropy easy axes, the latter make
different angles with respect to the field direction, and hence switch at
different values of the applied field.
This leads to a cluster-wise progressive switching of surface spins, as can be
seen in the structures of Fig.~\ref{struct_T_D10_j0_K1} at the respective field
values $h=0.47,0.56,0.64,1.12$ at which no spins, 2 spins, 3 spins, and all
spins have switched.
%
%
For the surface anisotropy constant used in Fig.~\ref{struct_T_D10_j0_K1} the
core switches before the full switching of the surface.

\begin{figure}[h!]
\begin{center}
 \includegraphics[angle=-90,width=3cm]{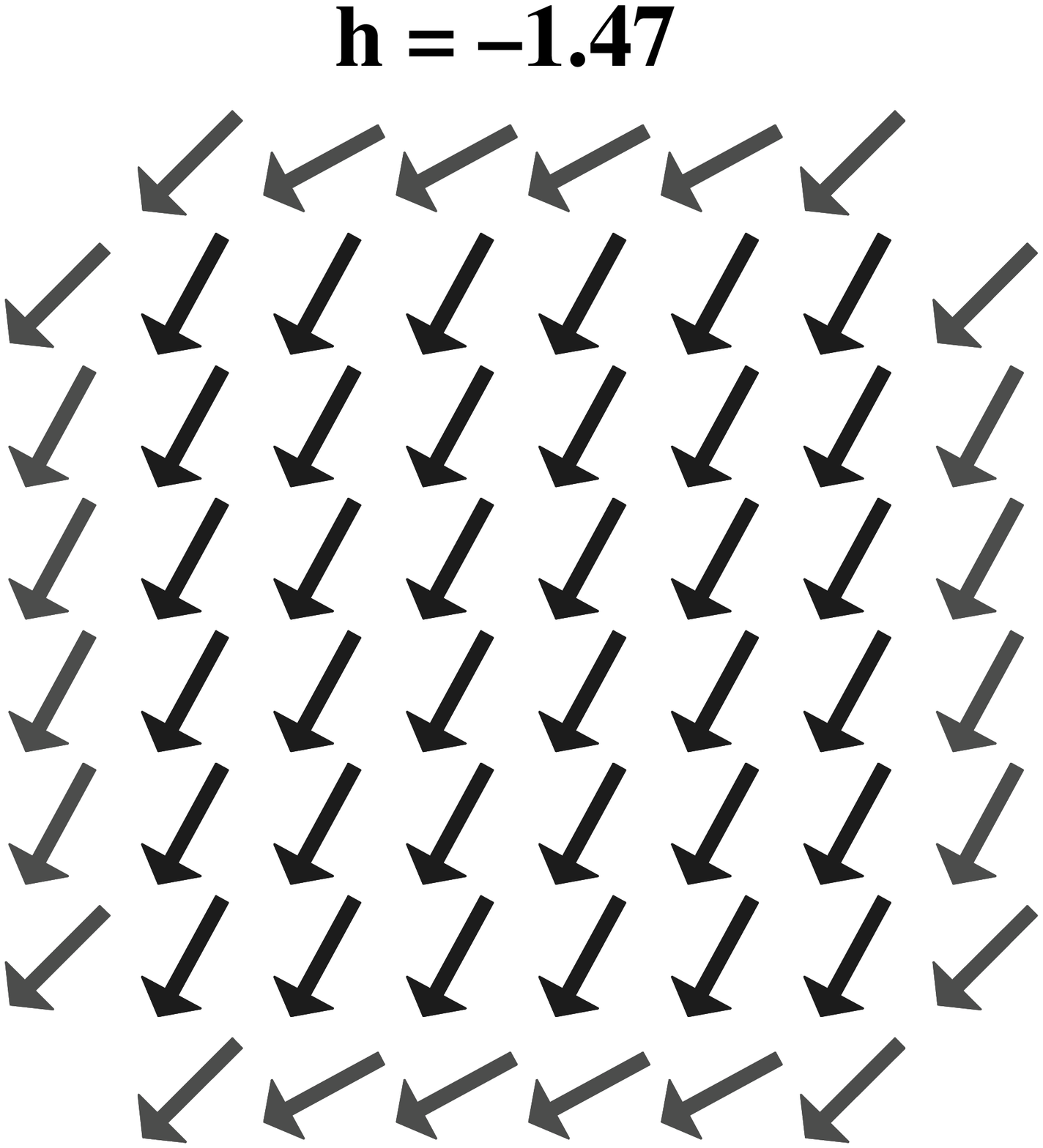}\hspace{0.5cm}
\includegraphics[angle=-90,width=3cm]{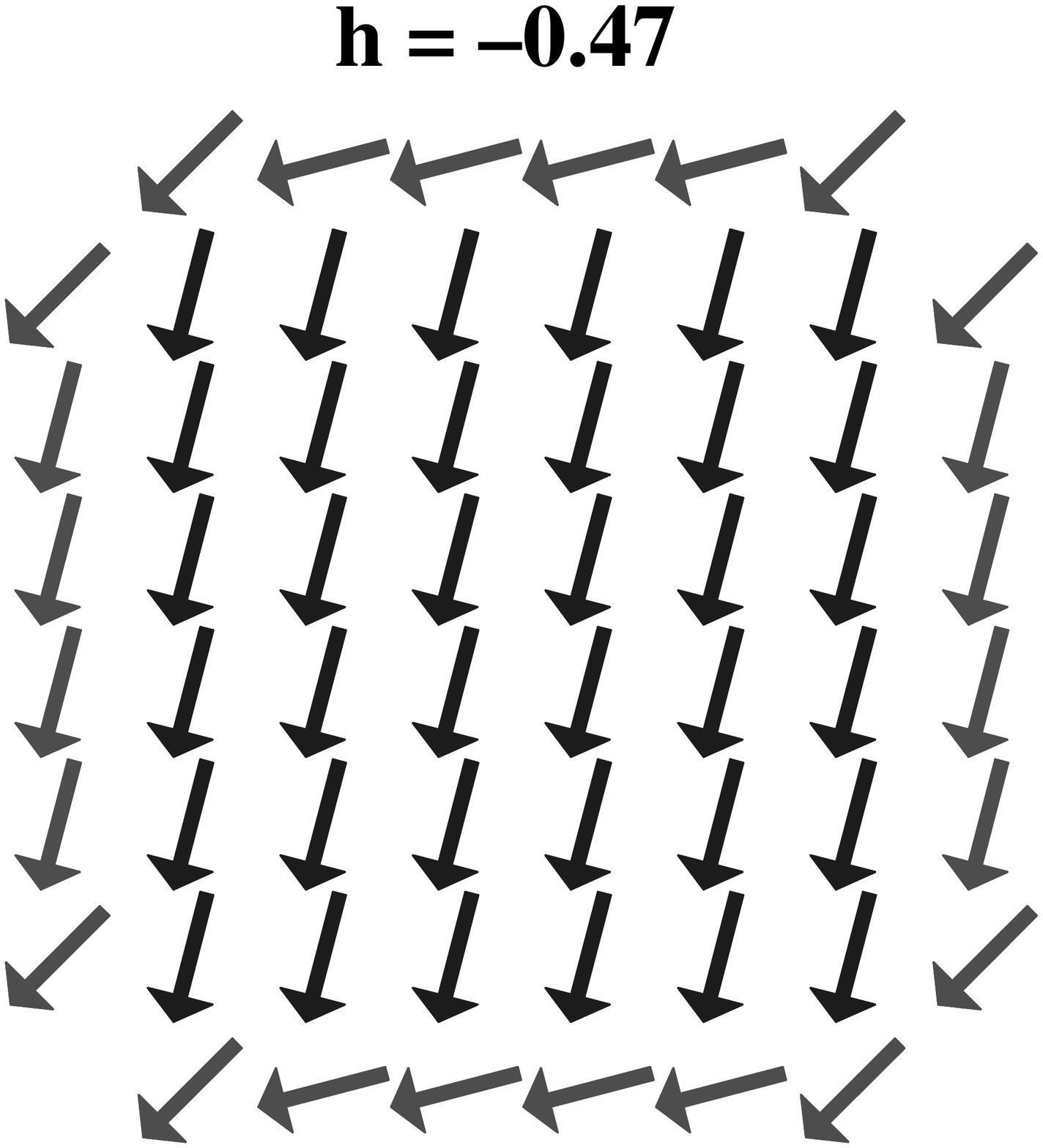}\hspace{0.5cm}
\includegraphics[angle=-90,width=3cm]{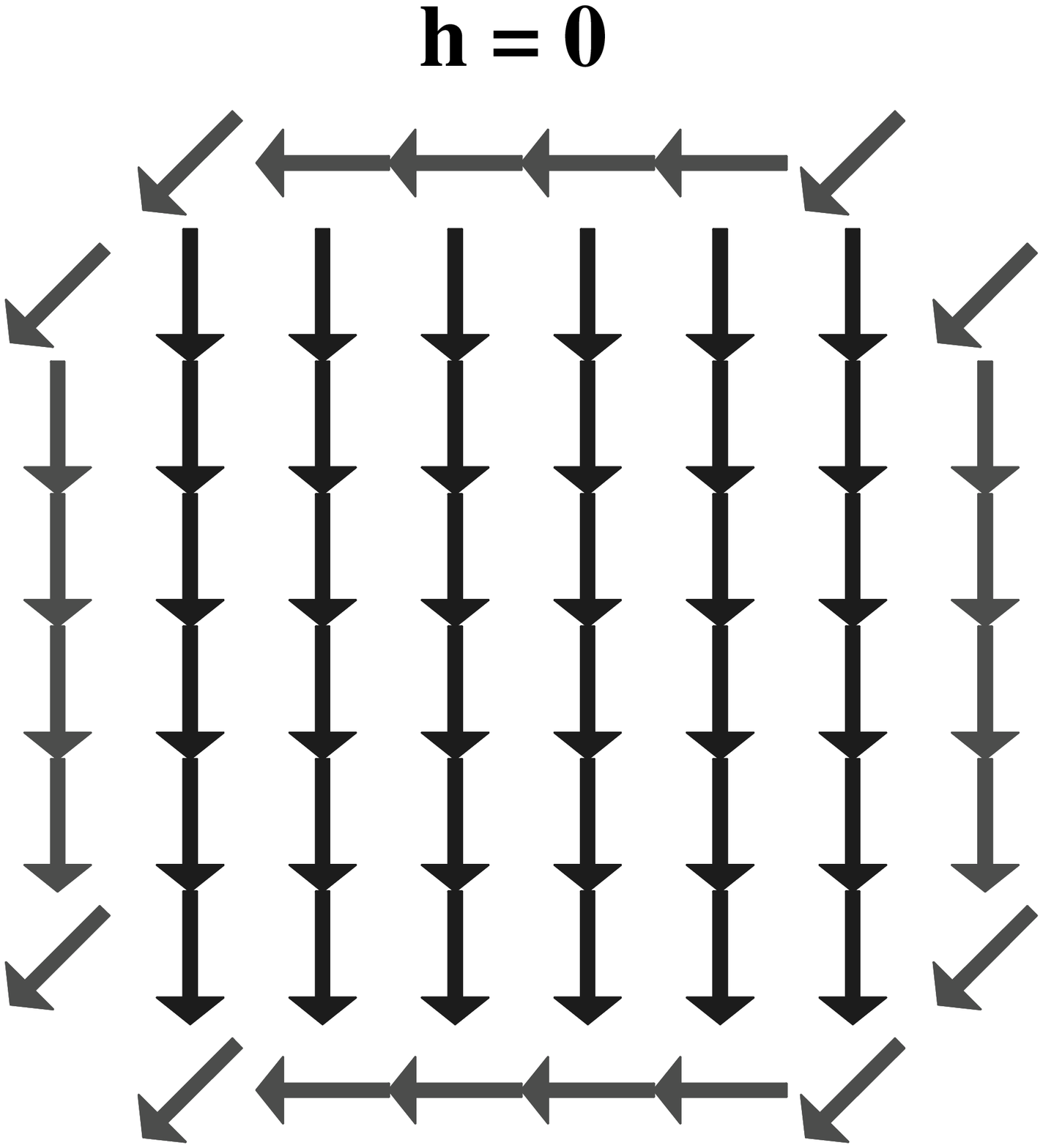}\hspace{0.5cm}
\includegraphics[angle=-90,width=3cm]{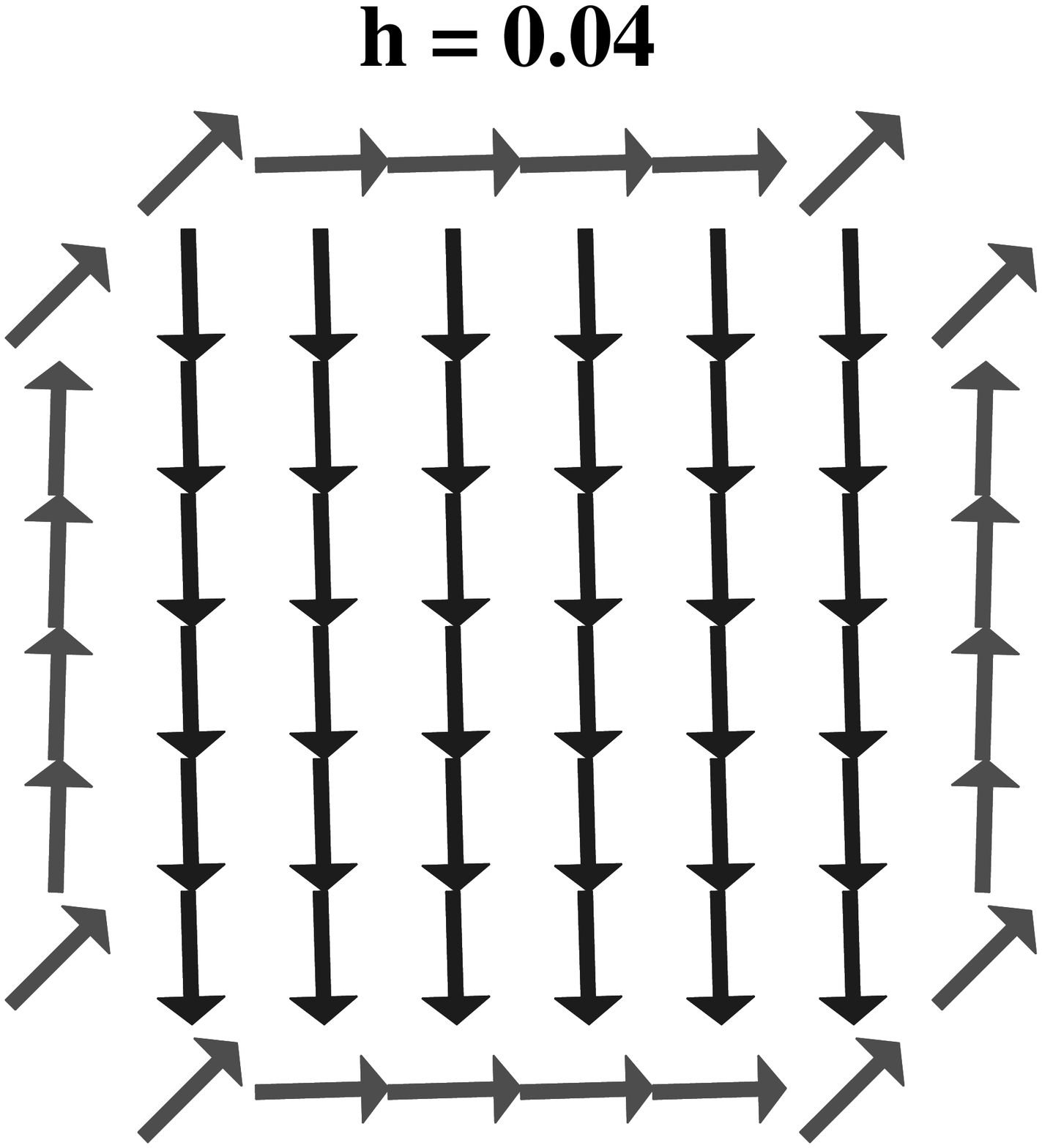}\\
\vspace{0.5cm}
\includegraphics[angle=-90,width=3cm]{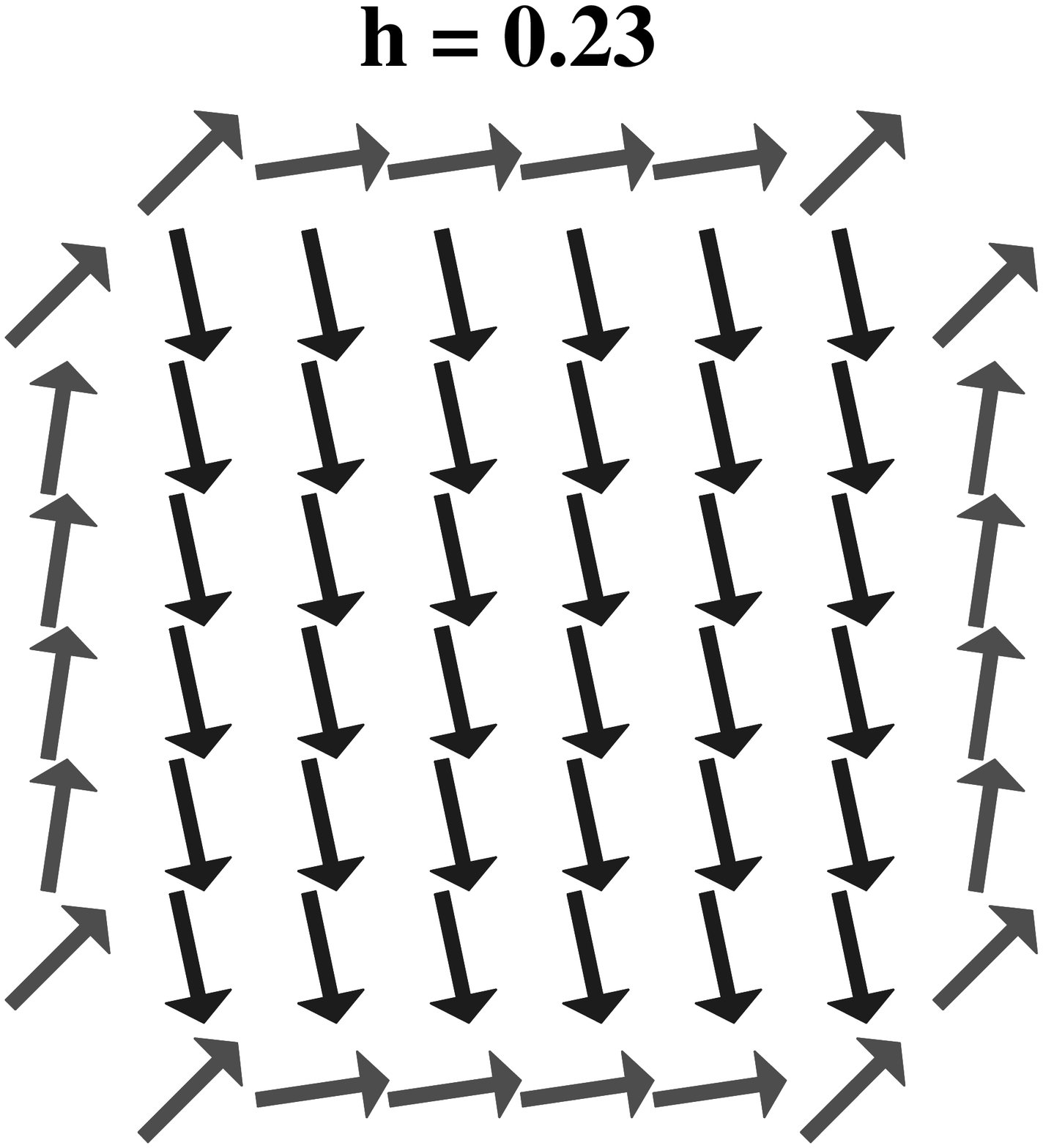}\hspace{0.5cm}
\includegraphics[angle=-90,width=3cm]{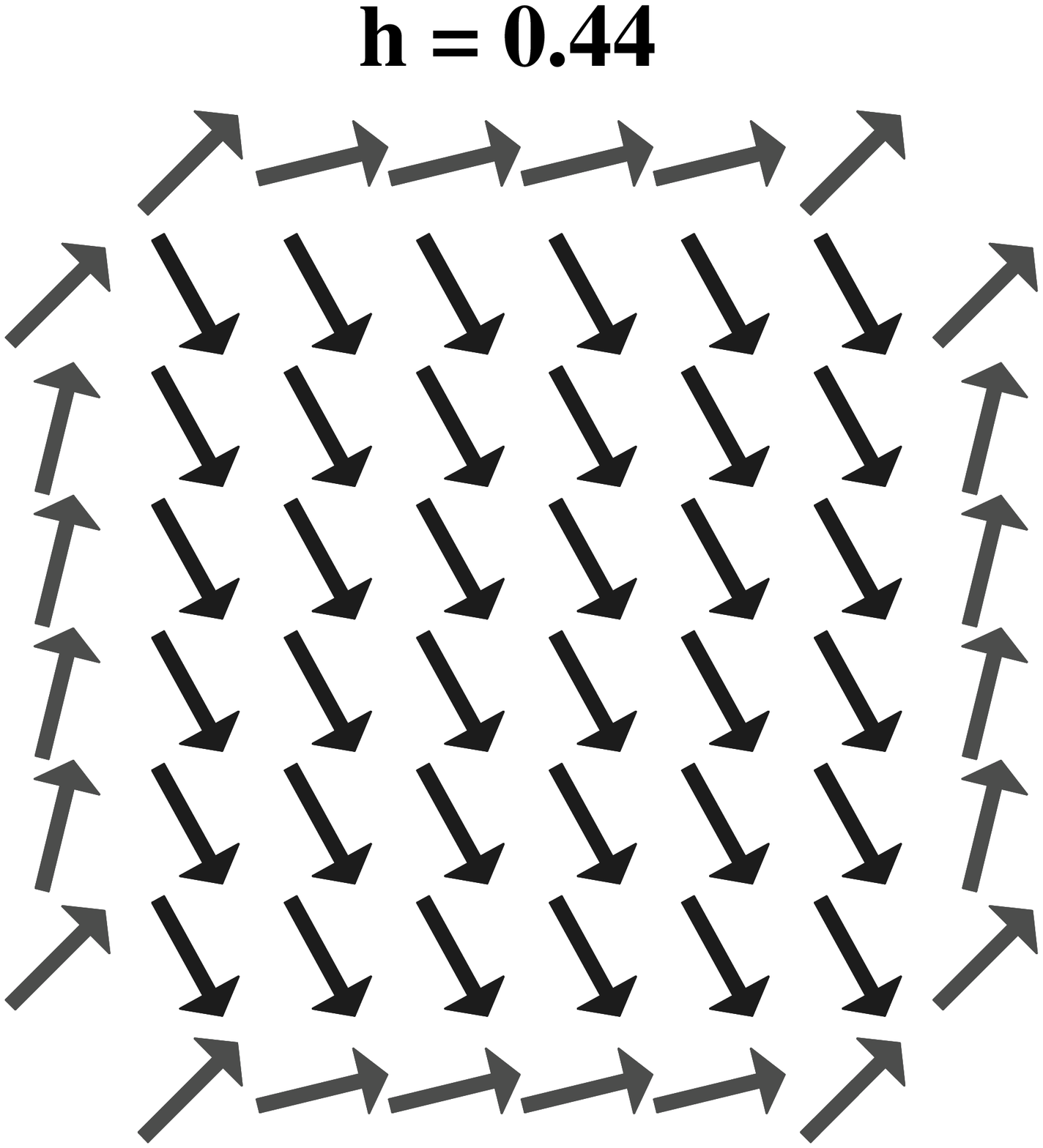}\hspace{0.5cm}
\includegraphics[angle=-90,width=3cm]{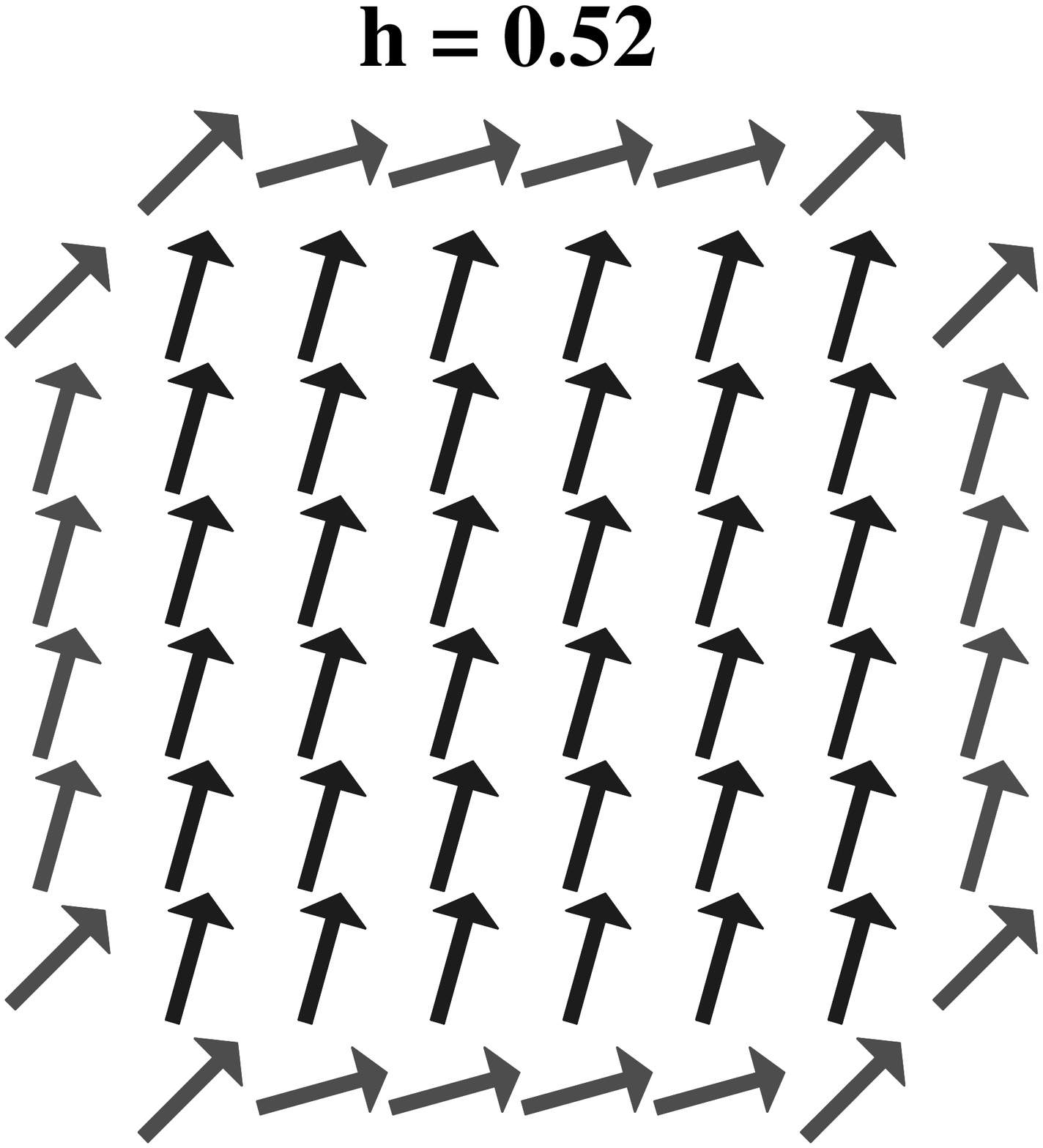}\hspace{0.5cm}
\includegraphics[angle=-90,width=3cm]{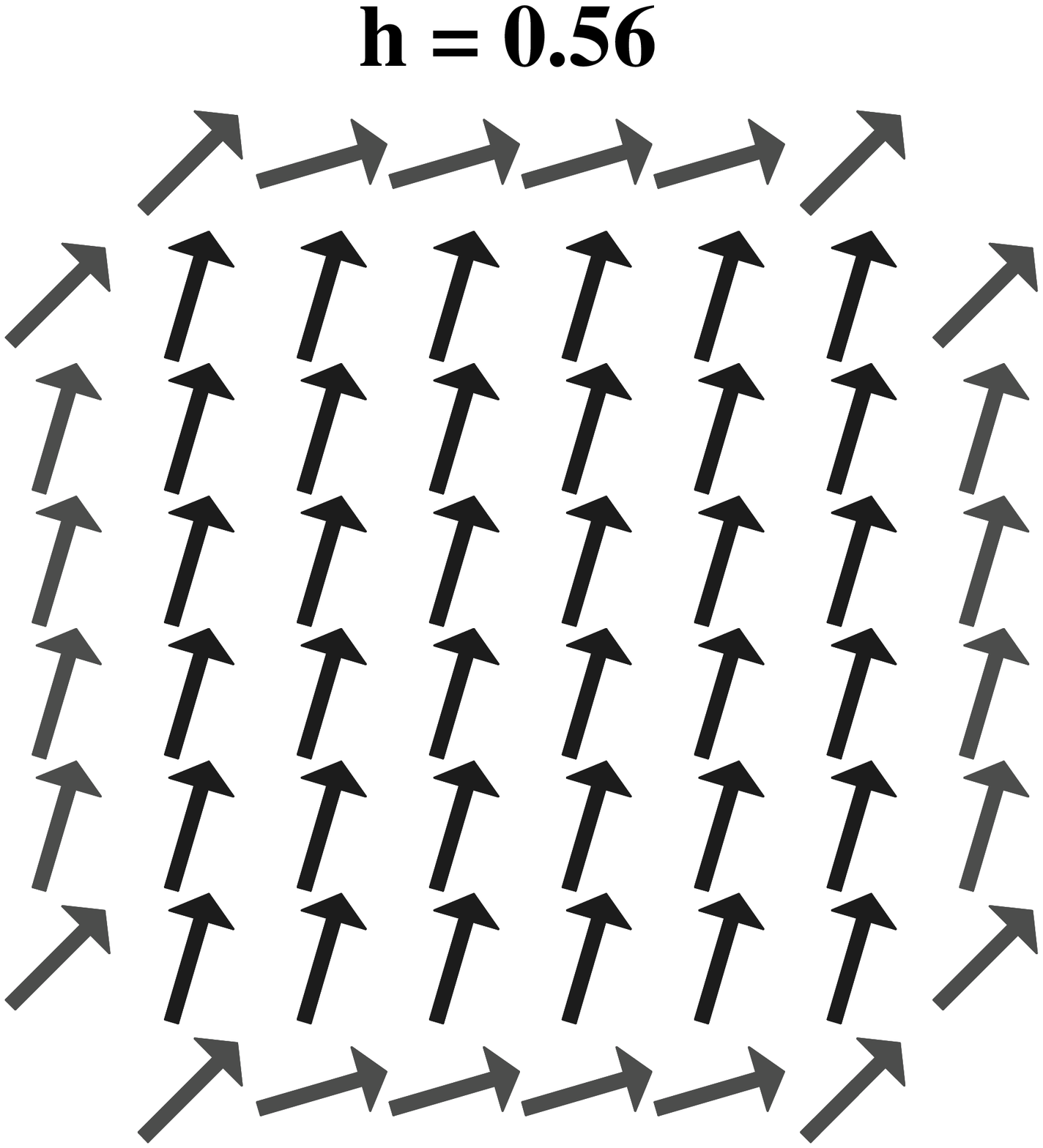}\\
\vspace{0.5cm}
\includegraphics[angle=-90,width=3cm]{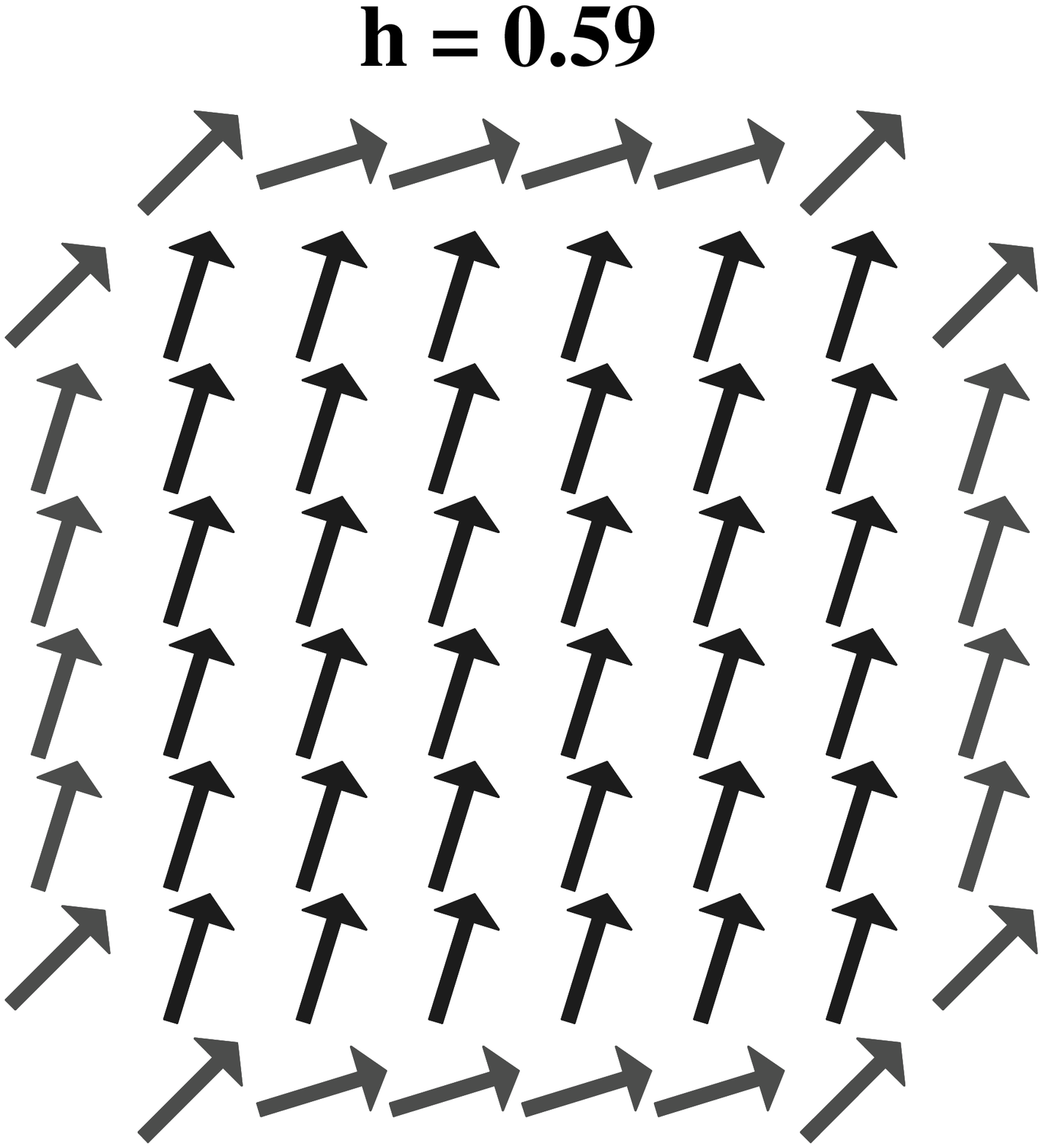}\hspace{0.5cm}
\includegraphics[angle=-90,width=3cm]{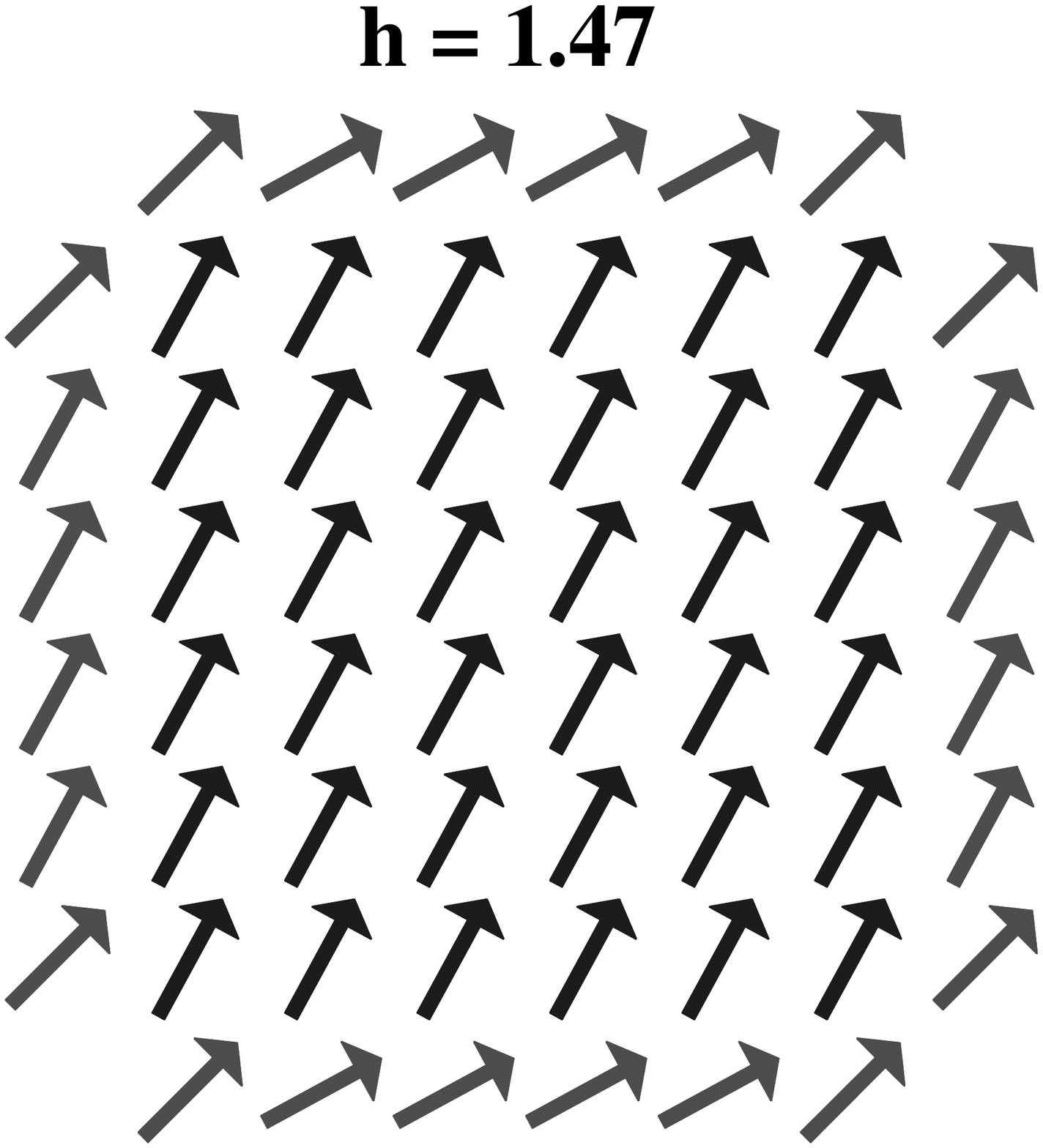}
\end{center}
\caption {\label{struct_N_D10_j0_K1}
Magnetic structures corresponding to the middle plane of the particle for the
NSA model.
The field values are indicated on top of the structures and correspond to the
ascending hysteresis-loop branch of Fig.~\ref{T_N_j0_D10_Ks1} for NSA.}
\end{figure}

In the case of NSA, the situation is quite different.
The hysteresis loop exhibits a first jump at a very small field, and more
importantly this jump corresponds to the switching of the entire surface.
This can be seen in Fig.~\ref{struct_N_D10_j0_K1} by examining the
structures at the field values $h=0$ and $h=0.04$.
For smaller particles with more spins on the boundary than in the core,
the first jump would also correspond to the switching of the particle's
magnetization.
%
%
Moreover, in Fig.~\ref{T_N_j0_D10_Ks1} there appear two jumps corresponding to
surface spins with two different local environments: the spins on the faces and
those located near the corners.
However, this difference in the switching field is just a numerical artefact,
since as will be shown shortly these two groups of spins jump at the same
field value, that is zero.
Indeed, a spin on the face has 5 nearest neighbors, so that the local energy
given by (\ref{NSA}) yields,
\begin{eqnarray}\label{energy_face}
e_i \equiv \frac{E_i}{2K_s} &=& -(s_x^2 + s_y^2 + s_z^2) +\frac{1}{2}s_z^2 -
{\bf h}\cdot{\bf s}_i \\ \nonumber
&\rightarrow& \frac{1}{2}\cos^2\theta -
\frac{h}{\sqrt{2}}\left[\sin\theta\cos\varphi+\cos\theta\right].
\end{eqnarray}
where $\theta,\varphi$ are the spherical coordinates of the spin ${\bf s}$.
The arrow in Eq.~(\ref{energy_face}) indicates that the irrelevant constant
$s_x^2+s_y^2+s_z^2=1$ is discarded.
So, in zero field the energy (\ref{energy_face}) has a minimum in the $x-y$
plane.
Similarly, a spin near the corner has three nearest neighbors (two on
the same plane and one on the adjacent plane), so that its energy reads
\begin{equation}\label{energy_corner}
e_i = -\frac{1}{2}(s_x^2 + s_y^2 + s_z^2) - {\bf h}\cdot{\bf s}_i
\rightarrow -
\frac{h}{\sqrt{2}}\left[\sin\theta\cos\phi+\cos\theta\right].
\end{equation}
whose minimum, in zero field, is along the diagonal of the cube [see
Eqs.~(\ref{En_greenfunct}, \ref{En_difference}) below and
Ref.~\cite{garkac03prl}].

As noted above, while the smallest switching field for a TSA spin is
$1/2$, here we see that the switching field of an NSA spin may assume very small
values.
Indeed, for spins on the faces and near the corners, it can be shown from
(\ref{energy_face}) and (\ref{energy_corner}), respectively, that the switching
field is zero.
For example, the magnetization direction which yields the lowest energy of an
NSA spin on the face is given by $\varphi=0$ and
\begin{eqnarray}\label{solution_nsa}
&&\cos\theta_{m} = \frac{1}{2}\left[A + \sqrt{1-2aA}\right], \nonumber\\
&&A \equiv a - \sqrt{1+a^2},\quad a\equiv \frac{h}{\sqrt{2}},
\end{eqnarray}
which when inserted in the determinant of the energy Hessian yields a zero
switching field. This is confirmed by the result of the numerical calculations
in Fig.~\ref{T_N_j0_D10_Ks1}, up to numerical precision.

In fact, the reason for such small switching fields for surface spins in the NSA
model is quite simple, and one actually does not need the above calculations to
convince onself.
Indeed, for spins on the face there is a rotation symmetry which makes all
directions in, e.g., the $x-y$ plane degenerate, see Eq.~(\ref{energy_face}).
So, when the field increases from the negative saturation value towards zero,
these spins select the direction given by the projection of the
field direction ${\bf e}_h$ on the $x-y$ plane.
As soon as the field points into the opposite direction, even a very small
value thereof is sufficient to make the spins on the face reverse their
direction.
The same arguments hold for spins near the corners, since these too have
full symmetry in the plane containing their easy direction.
\subsection{\label{subsec: Non0_exchange} Interacting case}
\begin{figure}[h!]
\begin{center}
\includegraphics[angle= -90, width= 14cm]{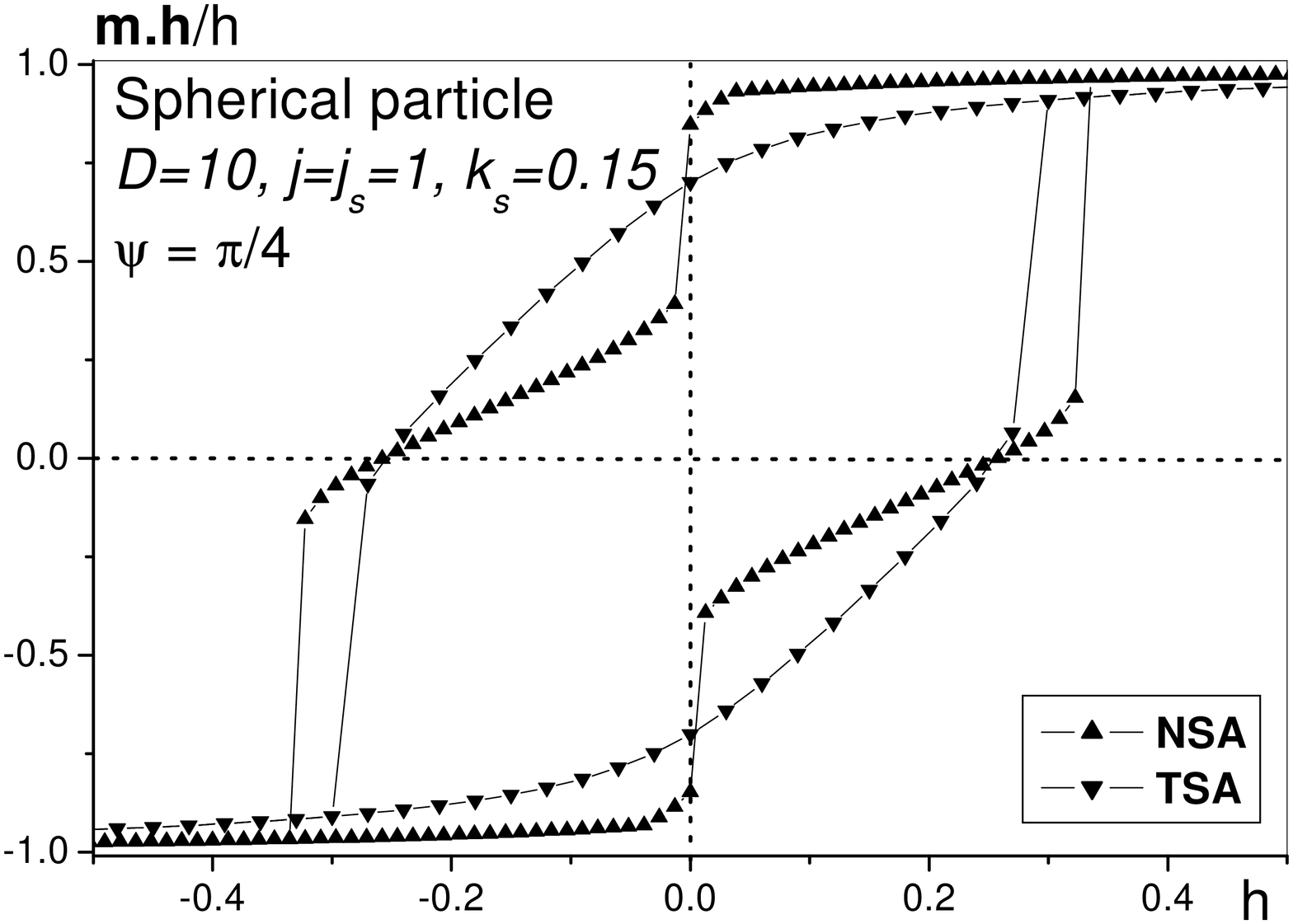}
\end{center}
\caption{\label{T_N_j1_D10_Ks015}
Hysteresis loops for a spherical particle with $D = 10, j = 1, k_s = 0.15$.
}
\end{figure}

\begin{figure}[h!]
\mbox{TSA}
\begin{center}
\includegraphics[angle=-90, width= 3cm]{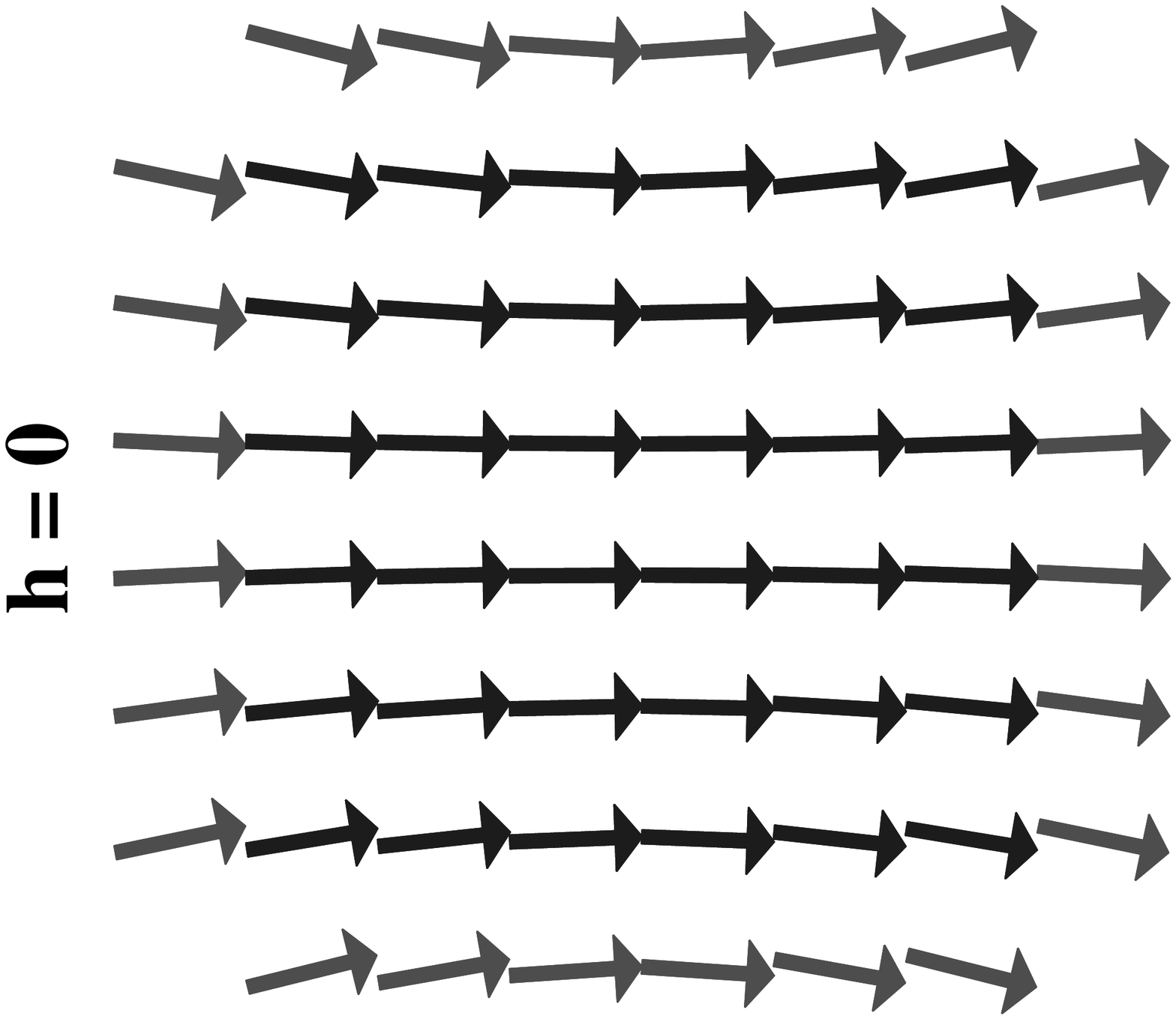}\hspace{0.5cm}
\includegraphics[angle=-90, width= 3cm]{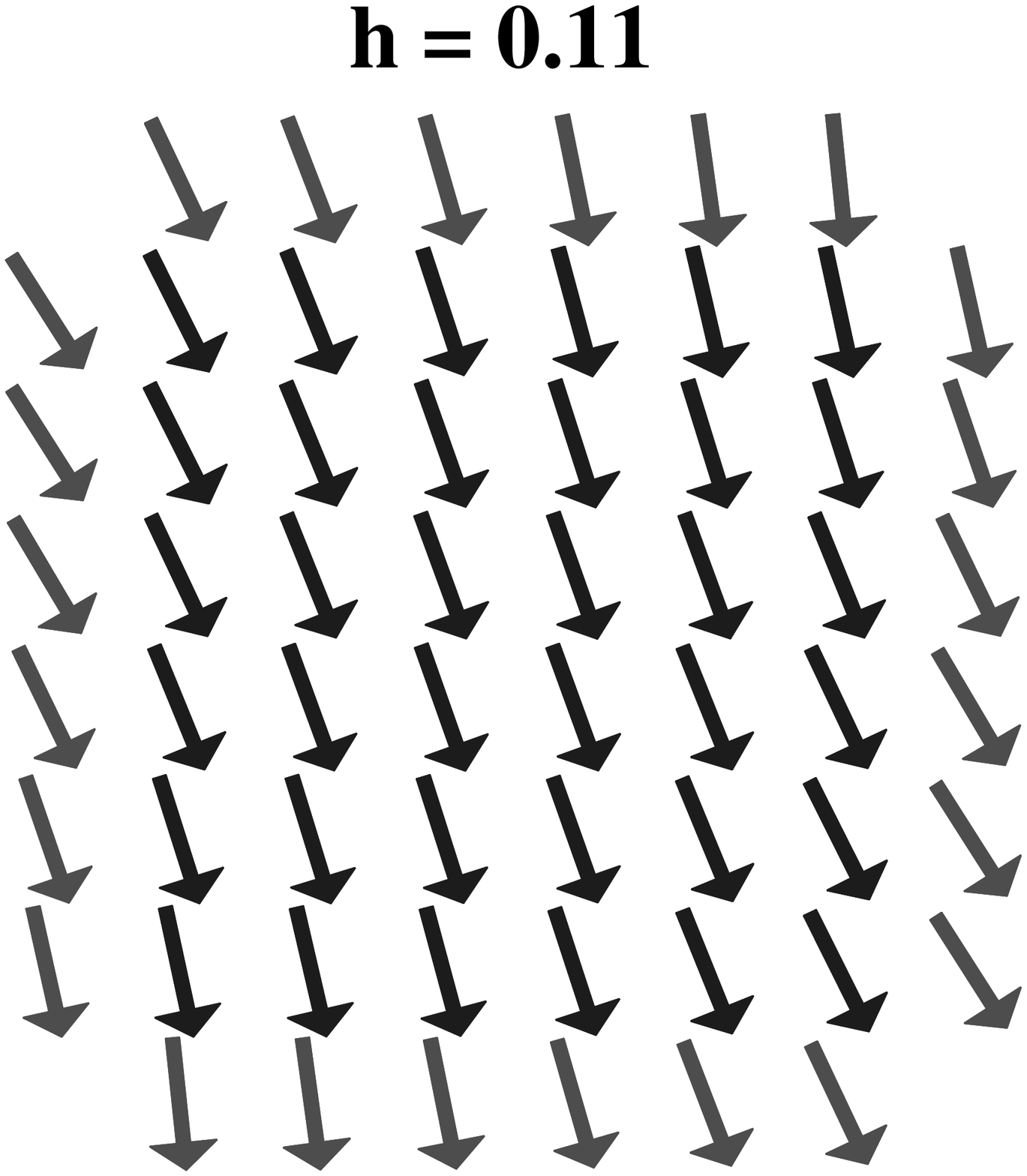}\hspace{0.5cm}
\includegraphics[angle=-90, width= 3cm]{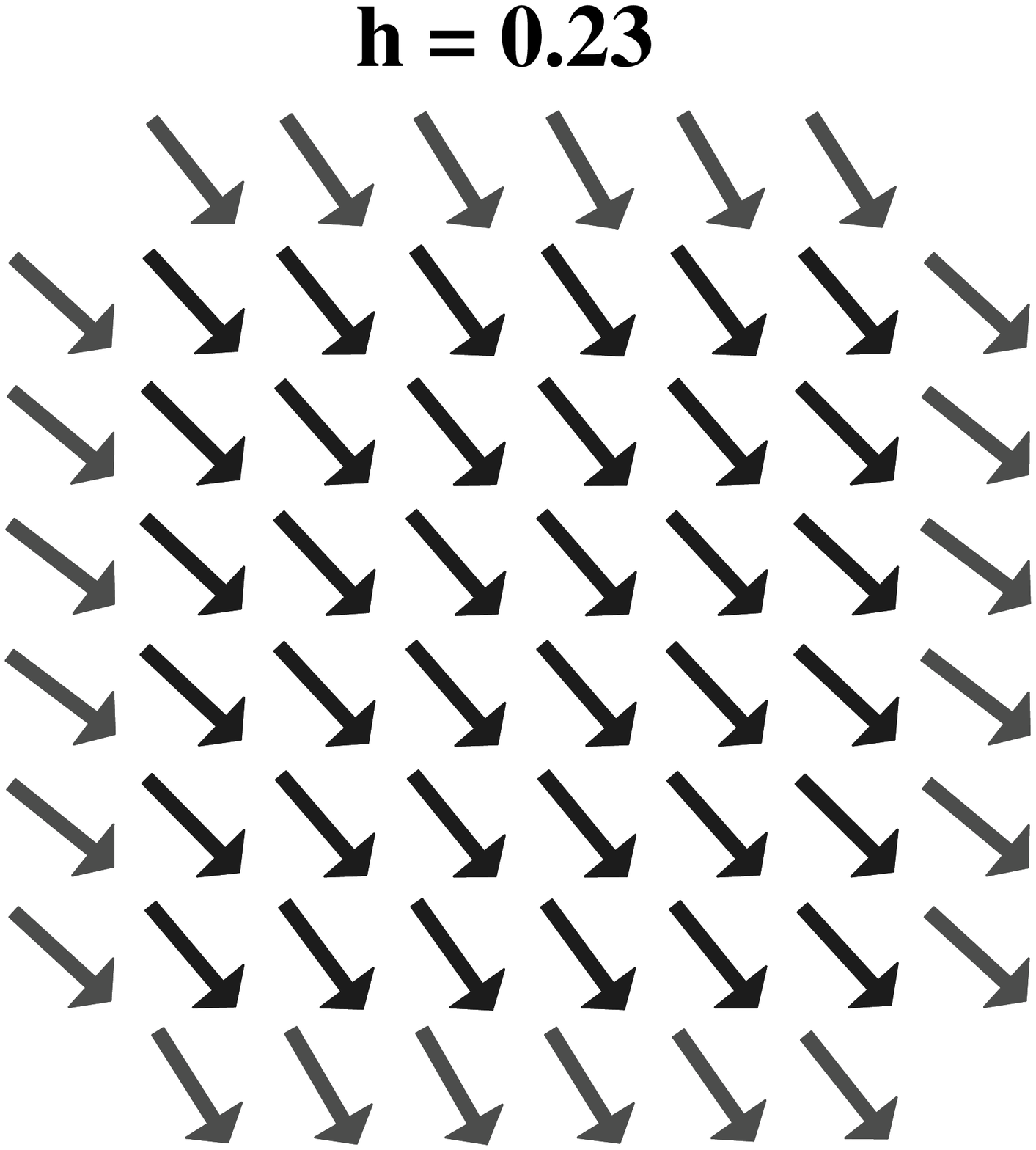}\hspace{0.5cm}
\includegraphics[angle=-90, width= 3cm]{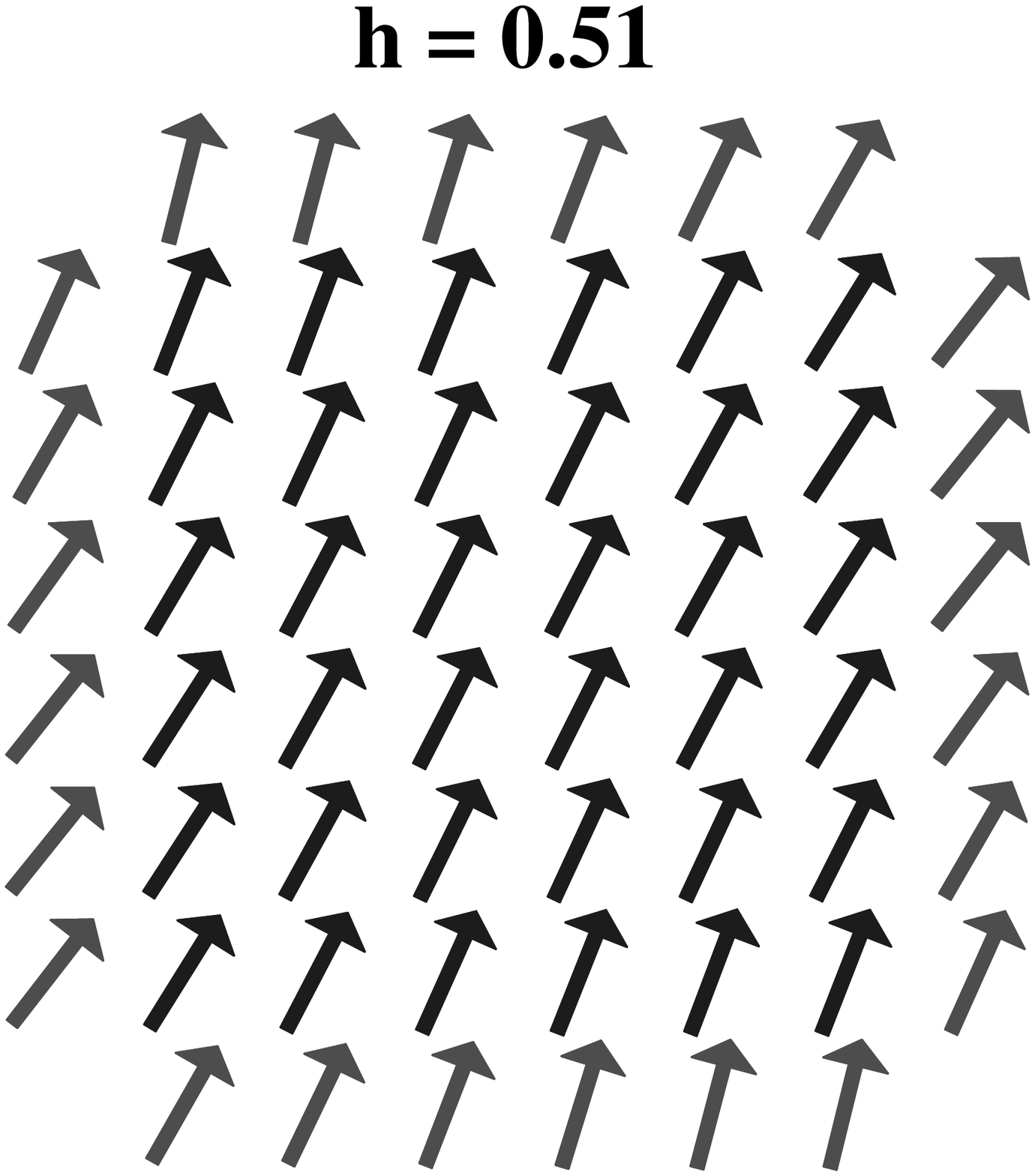}\\
\end{center}
\vspace{0.75cm}
\mbox{NSA}
\begin{center}
\includegraphics[angle=-90, width= 3cm]{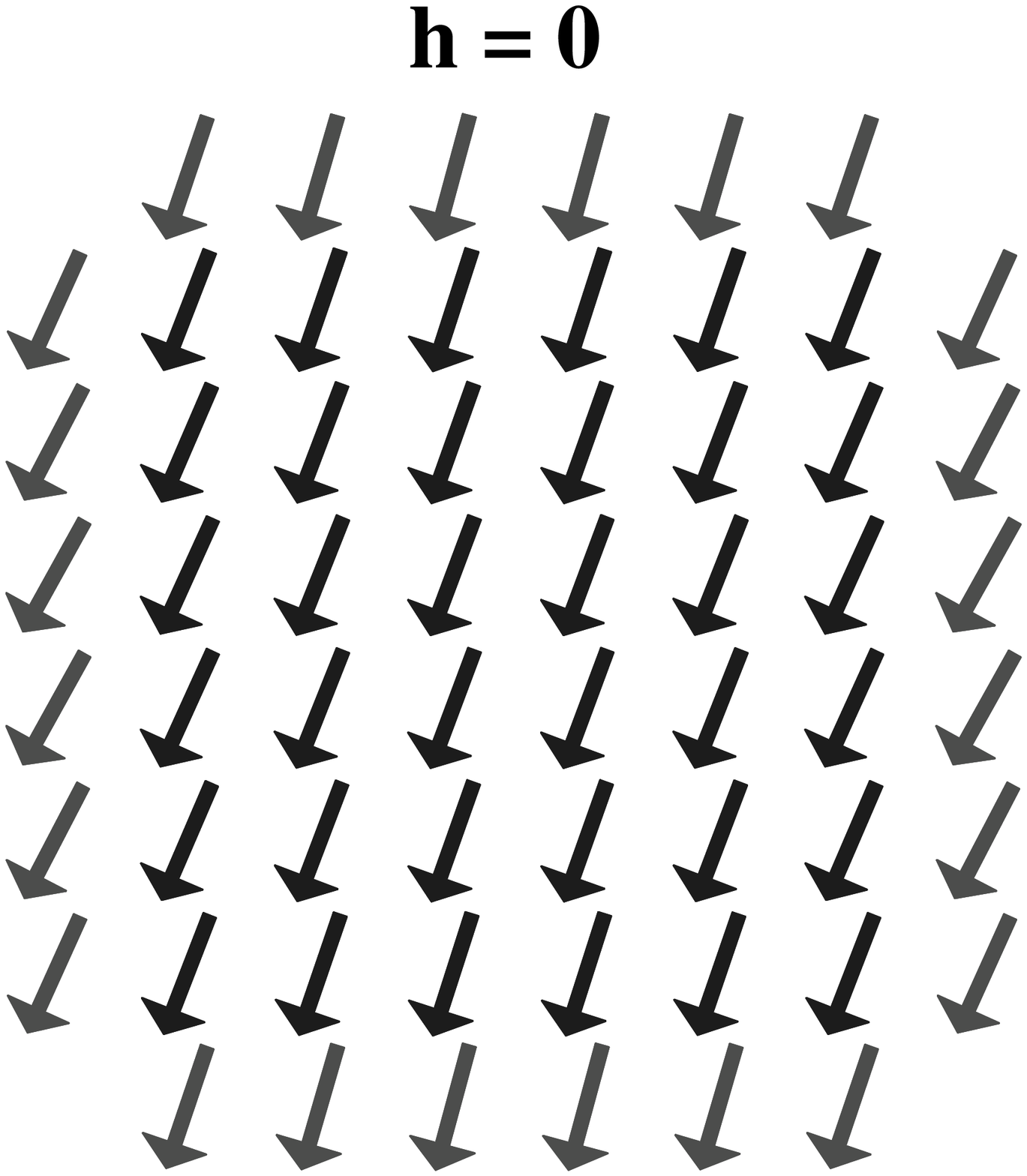}\hspace{0.5cm}
\includegraphics[angle=-90, width= 3cm]{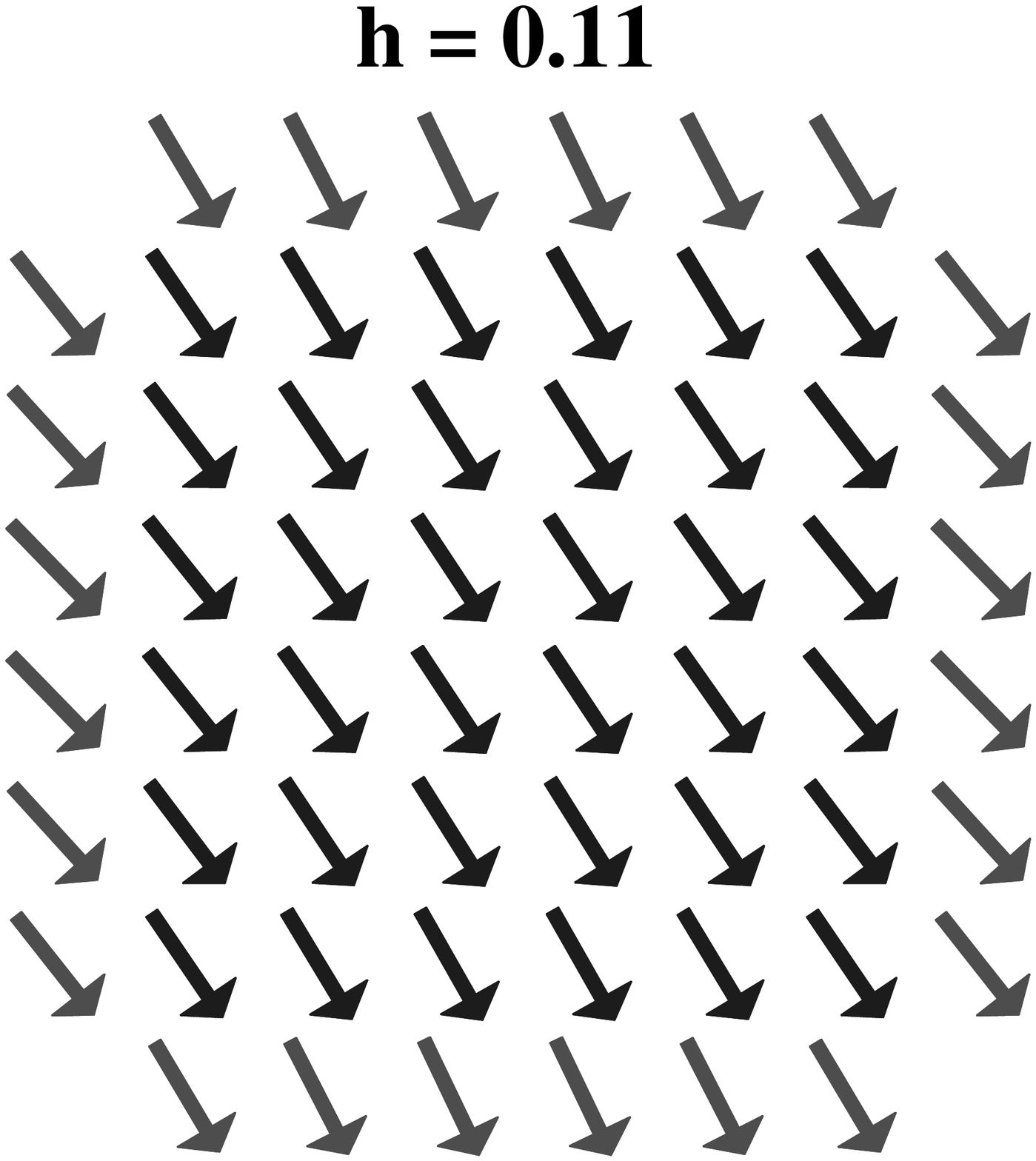}\hspace{0.5cm}
\includegraphics[angle=-90, width= 3cm]{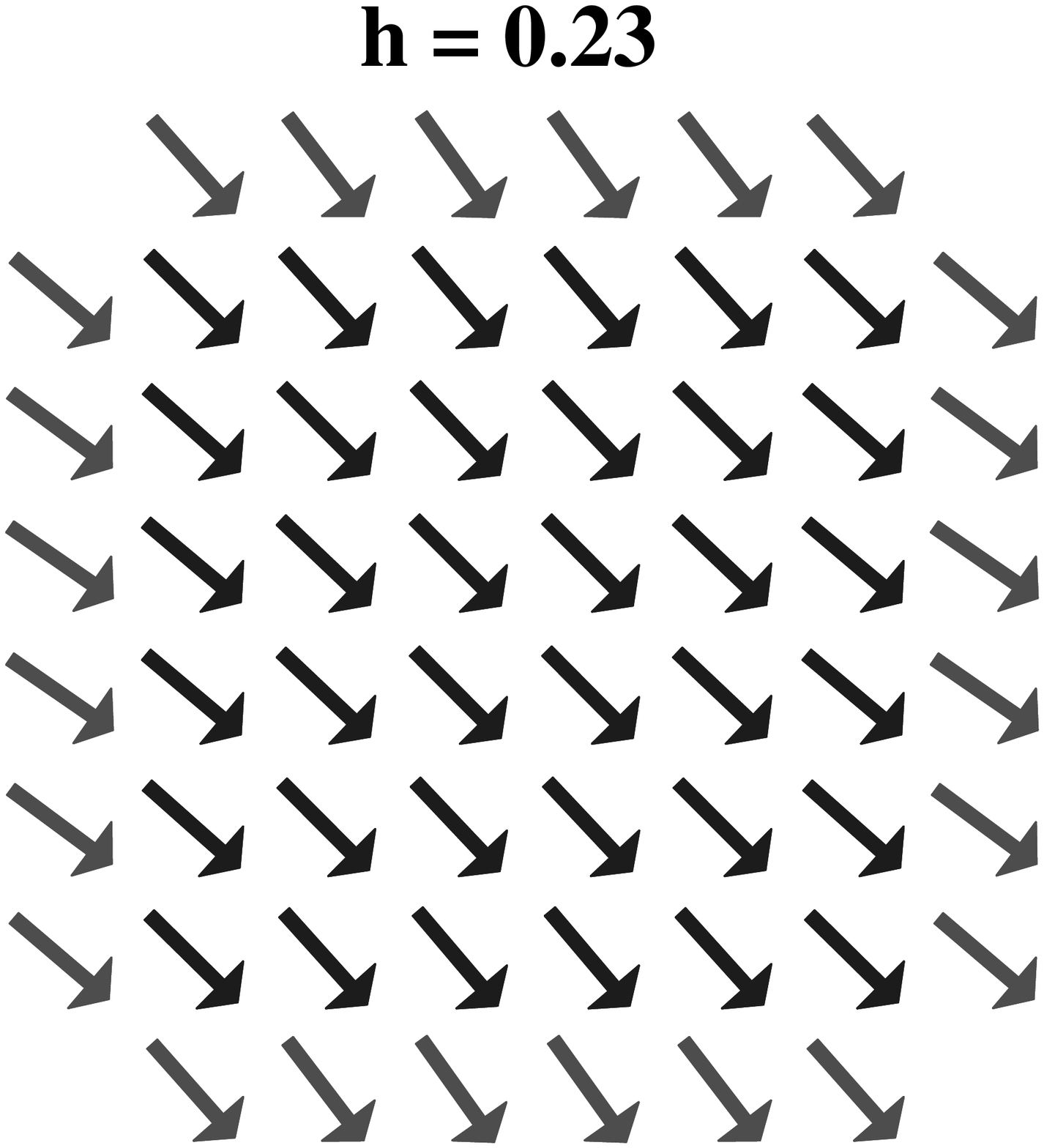}\hspace{0.5cm}
\includegraphics[angle=-90, width= 3cm]{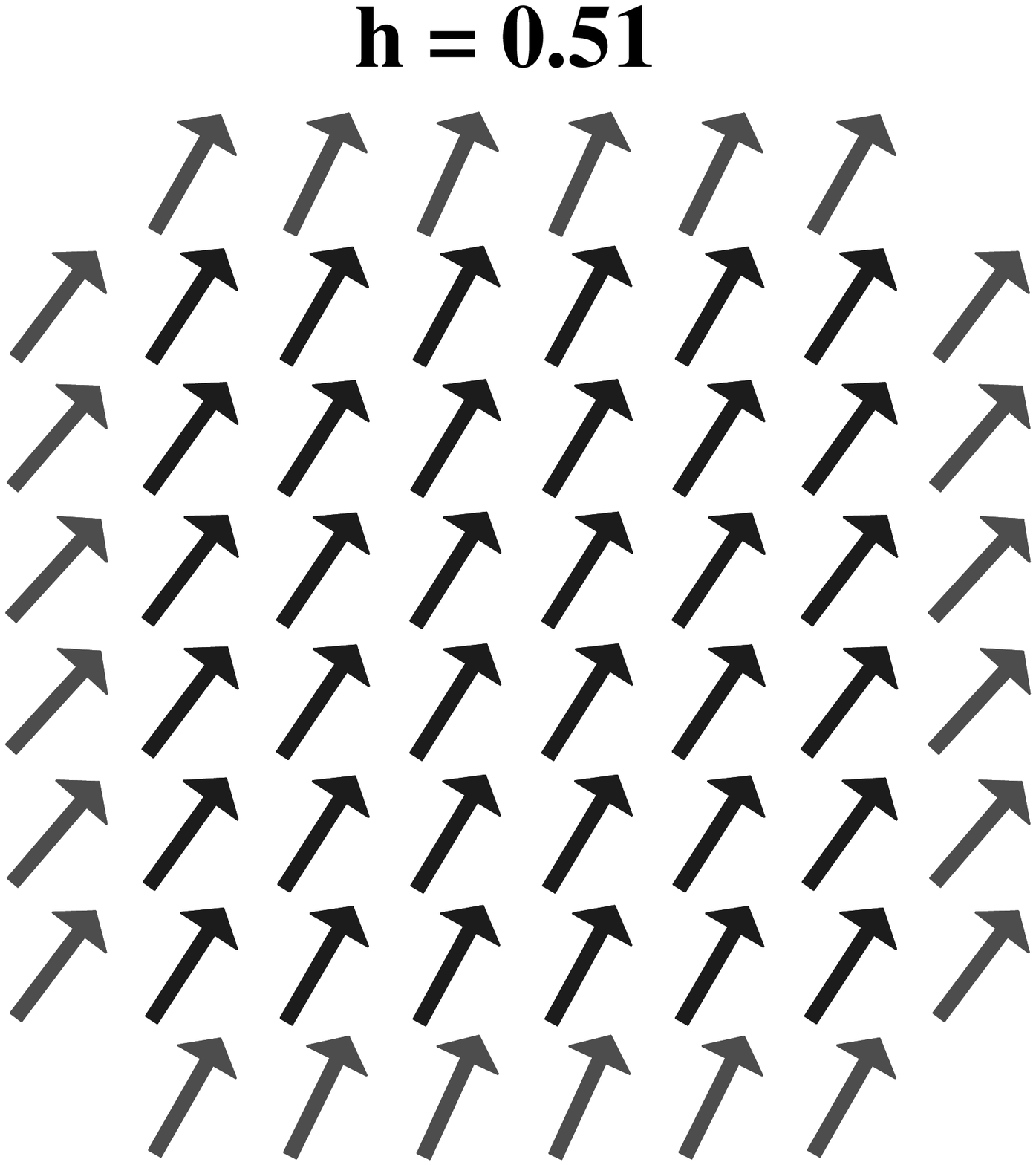}\\
\end{center}
\caption{\label{mag_struct_jnotzero}
Magnetic structures for the TSA and NSA particles with $D =
10, j = 1, k_s = 0.15$.
The field values on top of each structure correspond to the
ascending hysteresis-loop branches of Fig.~\ref{T_N_j1_D10_Ks015}.
}
\end{figure}

In order to investigate the surface influence on the core and vice-versa, we
consider the same particle as before but now with realistic non-vanishing
exchange coupling everywhere in the particle.
The results of the hysteresis loop and some of the corresponding spatial
distributions of the magnetization are shown in Figs.~\ref{T_N_j1_D10_Ks015} and
\ref{mag_struct_jnotzero}, respectively.
%
%
In this case of fully interacting spins, we see that as the field goes from
$0$ to $0.11$, the magnetization of a NSA particle experiences its first jump,
while that of the TSA particle only decreases (in absolute value) in a
progressive manner.
This is due to the fact that the angle of rotation between the direction at
$h=0$ and that at $h=0.11$, and thereby the difference in the projection of
magnetization along the field, is much larger for the NSA particle than for the
TSA one.
The reason is that at zero field, in the case of the TSA model, all core spins
and those surface spins located on the $x-y$ faces are directed along the easy
axis $z$, and hence make up a strong effective field that forces the spins on
the $x-z, y-z$ faces into nearly the same direction as in the core.
As to the NSA model, there are near-corner spins directed along the
diagonal of the cube, spins on the $x-y$ faces which lie in these faces, and
finally the more numerous core spins directed along their easy axis $z$.
This results in a competition which ends up with a net magnetization in a
direction inside the cube with its largest component along $z$.
These results suggest that the surface has a stronger effect in the
case of NSA model as compared to the TSA model. Indeed, in
Fig.~\ref{mag_struct_jnotzero} (NSA), for instance, we see that at zero field 
the core spins are deviated from their easy direction.
Moreover, in the non-interacting case, in the NSA
model, the surface jumps in a coherent way at a much lower field
than the core which also switches coherently.
In the interacting case, the surface imposes, via exchange coupling, its
switching at low fields on the core, and this results in an abrupt partial
coherent rotation of all spins, core and surface.
The core then proceeds to switch and drives the whole bunch of spins
at a higher field.

In \cite{garkac03prl} we studied the NSA model using both the numerical
solution of Landau-Lifshitz equation at zero temperature and the
analytical Green's function technique in the continuum limit, though preserving
the discreteness of the lattice by introducing a spin density function.
We showed that the contribution of the NSA to the energy of a spherical particle,
in the absence of core anisotropy and applied field, reads,
\begin{equation}\label{En_greenfunct}
{\mathcal E}=\kappa\frac{K_s^2{\mathcal N}}{J_0}(m_x^4+m_y^4+m_z^4),
\end{equation}
where $m_\alpha,\alpha=x,y,z$, are the components of the net magnetization of
the particle, $J_0=zJ$ is the Fourier transform of the exchange coupling
$J_{ij}$, $z=6$ the coordination number in the core, and $\kappa$ a
surface integral.
Computing the energy difference between two major orientations of ${\bf m}$
leads to
\begin{eqnarray}\label{En_difference}
\frac{J_0}{{\mathcal N}K_s^2}\left(E_{001}-E_{111}\right)&=&
\left\{
\begin{array}{ll}
4.2824, & \mbox{NSA}\\
0.2816, & \mbox{TSA}
\end{array}
\right. \\ \nonumber
\frac{J_0}{{\mathcal N}K_s^2}\left(E_{011}-E_{111}\right)&=&
\left\{
\begin{array}{ll}
1.0706, & \mbox{NSA}\\
0.0704, & \mbox{TSA}.
\end{array}
\right.
\end{eqnarray}
This indeed confirms what we said above, namely that NSA has a stronger effect
than TSA.
%
\begin{figure}[t]
\begin{center}
\includegraphics[angle=-90, width=14cm]{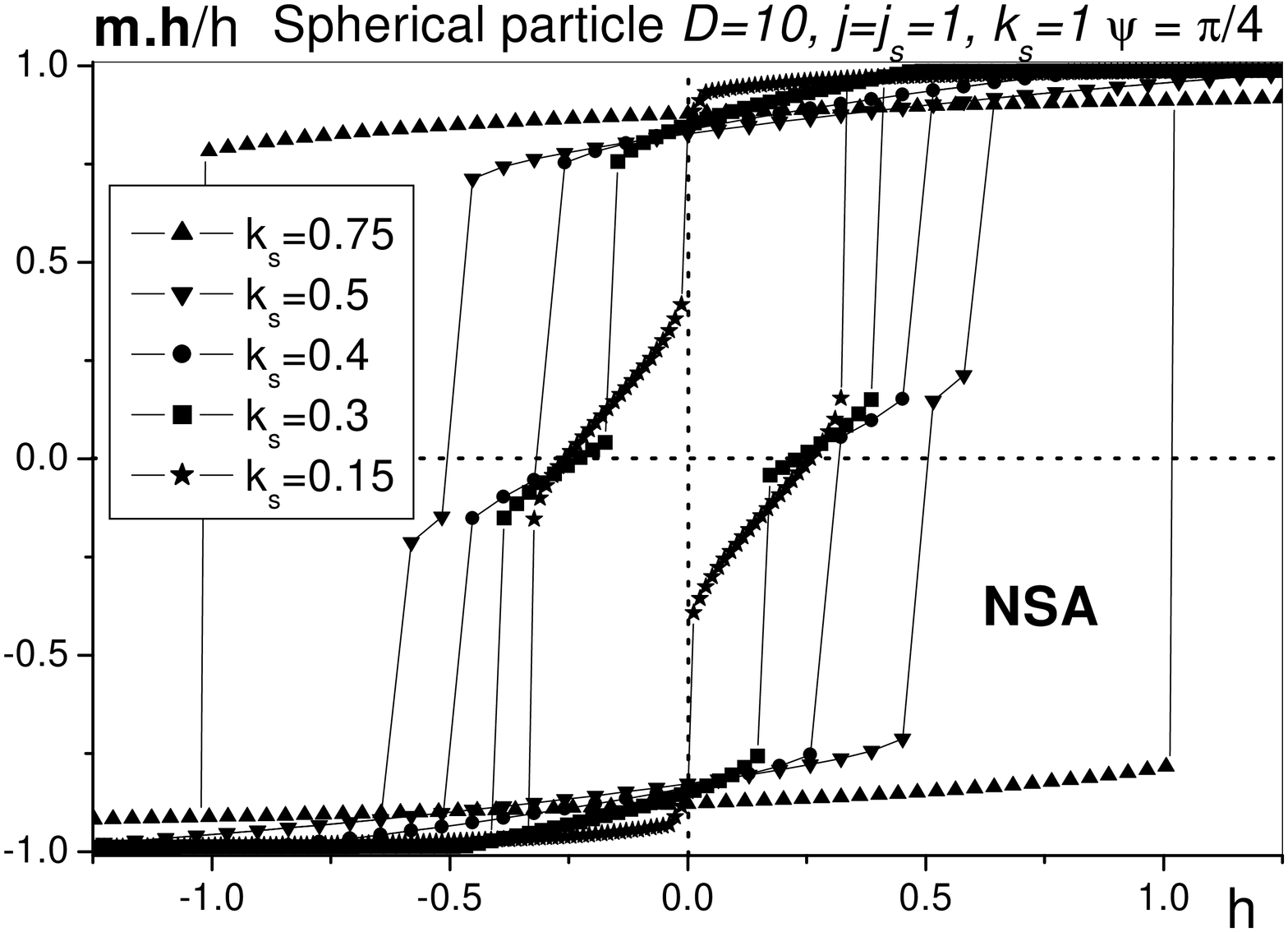}
\end{center}
\caption{\label{nsa_j1_D10_vsKs}Hysteresis loop for an NSA particle with
$j=1,D=10$ and variable $k_s$.}
\end{figure}
%

In Fig.~\ref{nsa_j1_D10_vsKs} we plot the hysteresis loop as obtained from the
NSA model for different values of the surface anisotropy constant $k_s$.
These results demonstrate that actually the amplitude of the abrupt coherent
rotation and the field at which it occurs increase with $k_s$.
Moreover, the length of the subsequent ``shoulder" corresponding to a smooth
rotation of the whole spins and that precedes the magnetization switching
decreases with $k_s$ and vanishes at some higher value of $k_s$ ($0.75$ here).
This means that as $k_s$ becomes large, the amplitude (or angle) of the abrupt
coherent rotation, induced by surface anisotropy, becomes so large that this
rotation coincides with the switching of the net magnetization and thus no other
steps are observed.

In summary, surface anisotropy in the TSA model induces several jumps in the
hysteresis loop because of the cluster-wise switching of
spins~\cite{kacdim02prb}.
This is due to the single-site nature of
the anisotropy in this model, and to the radial direction of surface anisotropy
axes, and thence to the distribution of angles that these axes make with 
the field direction.
In the NSA model, the existence of several jumps in the hysteresis loop does
not require high values of the surface anisotropy constant [see Fig.~\ref{T_N_j1_D10_Ks015}],
and is due to successive coherent partial rotations of all spins in the
particle.
For large values of the  surface anisotropy constant, there is only one big
coherent rotation which coincides with the net magnetization switching.

As in Ref.~\cite{kacdim02prb} where the switching field was also studied as
a function of the surface anisotropy constant $k_s$, we have performed the same
calculation here for the NSA model, for the sake of comaprison.
%
\begin{figure}[h!]
\begin{center}
\includegraphics[angle= -90, width= 14cm]{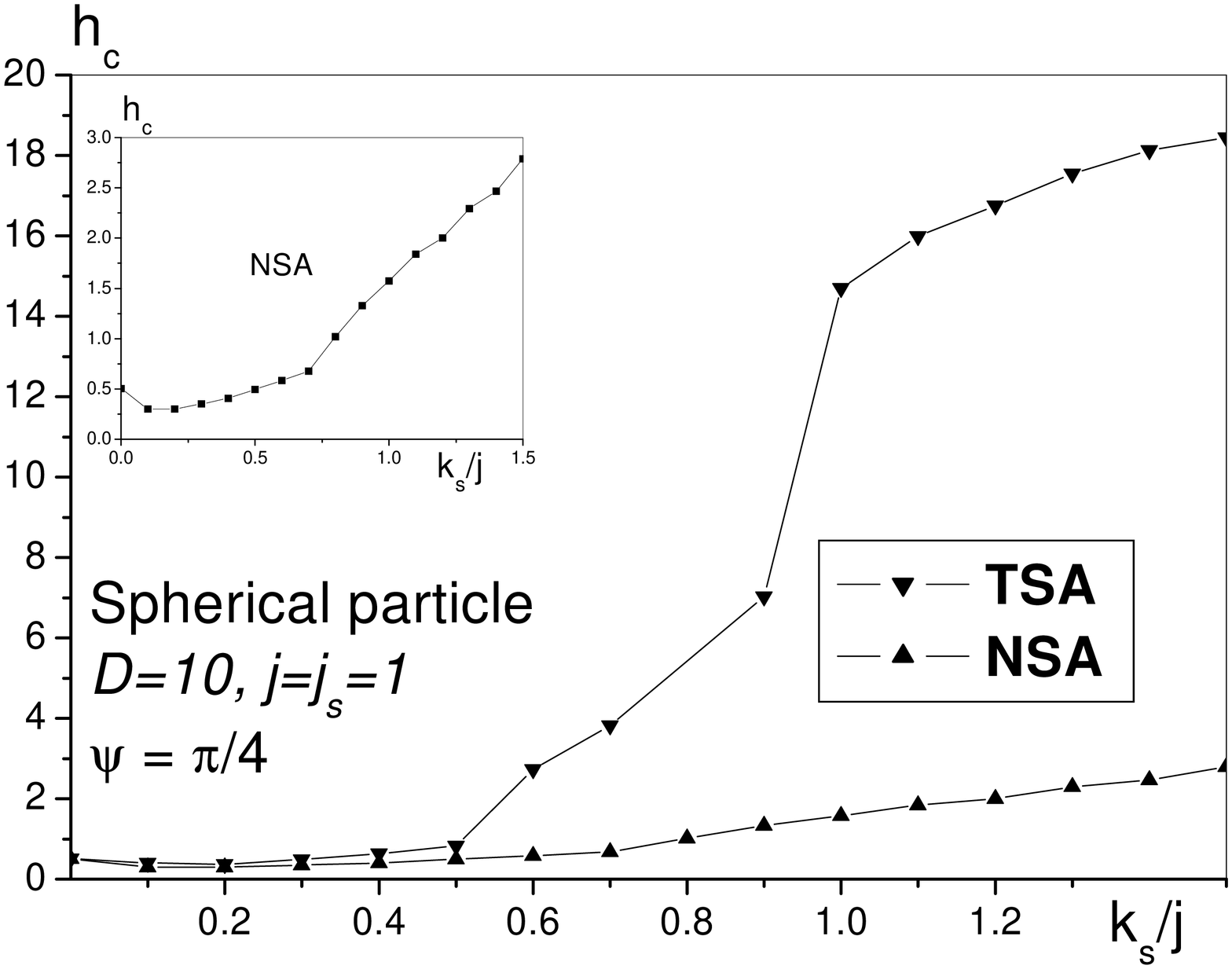}
\end{center}
\caption{\label{hcvsks_T_N}
Switching field against surface anisotropy for a spherical particle with
$D = 10, j = 1,\psi=45, $ for both TSA and NSA models. The inset gives
$h_c(k_s/j)$ for NSA alone.}
\end{figure}
%
The results are presented in Fig.~\ref{hcvsks_T_N}.
For the TSA model, there is a sudden increase in $h_c(k_s/j)$ occurring at a
value of $k_s/j$ of order one.
This was interpreted in \cite{kacdim02prb} as a critical value marking a
departure from the Stoner-Wohlfarth model, i.e, a transition from the
coherent-rotation regime, where the hysteresis loop exhibits only one jump, into
a regime of cluster-wise magnetization switching, where several jumps appear in
the hysteresis loop.
In the case of NSA, $h_c$ assumes much lower values than in the
TSA model as already discussed earlier, and increases in a more progressive way.
On the other hand, it is apparent from the inset in Fig.~\ref{hcvsks_T_N} that
there is in fact some kind of transition between different regimes, though
with a lower ``critical value" for $k_s/j$ than for TSA model.
\section{\label{sec:disc}Conclusion}
We have compared the effects of transverse surface anisotropy (TSA) and
N\'eel's surface anisotropy (NSA) on the hysteretic properties of a spherical
particle.
In the non-interacting case, i.e., with zero exchange coupling everywhere
inside the particle, the NSA induces more jumps than the TSA in the hysteresis
loop the first of which occurs at very low fields, and corresponds to the
switching of the entire surface.
This is due to the rotational symmetry which renders all directions in, e.g.,
the $x-y$ plane, degenerate.
Moreover, in the NSA model, the spin switching within the surface and within
the core is coherent.
In the interacting case with the same exchange coupling in the
whole particle, we still have qualitatively the same hysteresis loop as before,
but now the first jump in the NSA model occurs at a field slightly larger than
before, and more importantly corresponds to a coherent partial rotation of all
spins (core and surface).
More precisely, in this case the surface imposes its switching on the core,
and thereby the two jumps in the hysteresis loop, in fact, correspond to two
successive coherent rotations of the whole bunch of spins inside the particle.
This shows that, for the same constant, the surface anisotropy in the NSA model
has a stronger effect than in the TSA model.
Moreover, spin switching in TSA is cluster-wise, while that for
NSA is coherent but operates by successive coherent partial rotations.

Experimental results on individual cobalt clusters
[see the review \cite{wer01acp}] on the Stoner-Wohlfarth astroid, i.e., the
angular dependence of the switching field, together with the switching field
distribution and the non-switching probability, have confirmed the validity of
the Stoner-Wohlfarth and N\'eel-Brown models describing the magnetization
reversal by uniform rotation~\cite{weretal97prl} at zero and non-zero
temperature, respectively.
On the other hand, even though relatively large surface anisotropy constants
are obtained for cobalt particles \cite{jametal01prl}, the results are still
well fitted by the aforementioned models that are based on coherent rotation of
the net magnetization.
This is at variance with what one expects from the TSA model with large
anisotropy constant, as was demonstrated in \cite{kacdim02prb} and
reviewed here.
Therefore, in addition to the fact the NSA model is intuitively more realistic,
these experimental indications together with the present calculations give a
further hint that the surface anisotropy in nanomagnets is most probably of the
NSA kind.
However, a more convincing check would only come from direct
measurements of the hysteresis loop of an individual nanoparticle, were they
to become possible.
\section{Acknowledgments}
H. Mahboub thanks the Laboratoire de Magn\'etisme et d'Optique for financial
support and for the hospitality extended to her during her training 1 December
2002 - 31 July 2003.

\end{document}